\documentclass{aa}  

\usepackage{graphicx}
\usepackage{txfonts}
\usepackage{natbib}
\bibpunct{(}{)}{;}{f}{}{,} % to follow the A&A style

\newcommand{\kms}{\,$\mathrm{km\, s^{-1}}$}
\newcommand{\ms}{\,$\mathrm{m\, s^{-1}}$}
\newcommand{\Teff}{\ensuremath{T_{\mathrm{eff}}}}
\newcommand{\logg}{\ensuremath{\log g}}

\newcommand{\Mj}{\ensuremath{M_{\mathrm{Jup}}}}
\newcommand{\SM}{\mbox{$M_{\odot}$}}
\newcommand{\JM}{\mbox{$M_{J}$}}

\begin{document}

  \title{Search for giant planets in M67 IV: survey results
  \thanks{Based on observations collected at the ESO 3.6m telescope (La Silla), 
   at the 1.93m telescope of the Observatoire de Haute-Provence (OHP, France), 
   at the Hobby Eberly Telescope (HET, Texas), at the Telescopio Nazionale Galileo (TNG, La Palma) 
   and at the Euler Swiss Telescope (La Silla).}}

  \author{A. Brucalassi\inst{1,2,3} \and J. Koppenhoefer\inst{1,2} \and R. Saglia\inst{1,2} \and  L. Pasquini\inst{3} \and M.T. Ruiz\inst{4} \and P. Bonifacio\inst{5} 
  \and L. R. Bedin\inst{6} \and M. Libralato\inst{6} \and K. Biazzo\inst{7} \and C. Melo \inst{8}\and 
  C. Lovis\inst{9} \and S. Randich\inst{10}   }

   \institute{Max-Planck f\"ur extraterrestrische Physik, Garching bei M\"unchen, Germany
   \and University Observatory Munich, Ludwig Maximillian Universitaet, Scheinerstrasse 1, 81679 Munich, Germany
   \and ESO -- European Southern Observatory, Karl-Schwarzschild-Strasse 2, 85748 Garching bei M\"unchen, Germany 
   \and Astronomy Department, Universidad de Chile, Santiago, Chile 
   \and GEPI, Observatoire de Paris, CNRS, Univ. Paris Diderot, Place Jules Janssen 92190 Meudon, France
   \and Istituto Nazionale di Astrofisica, Osservatorio Astronomico di Padova, Padova, Italy
   \and Istituto Nazionale di Astrofisica, Osservatorio Astrofisico di Catania, Catania, Italy
   \and ESO -- European Southern Observatory, Santiago, Chile
   \and Observatoire de Geneve, Sauverny, CH
   \and Istituto Nazionale di Astrofisica, Osservatorio Astrofisico di Arcetri, Firenze, Italy
   }          
             
  % \date{Received September 15, 1996; accepted March 16, 1997}
    \date{Received  / Accepted }

% \abstract{}{}{}{}{} 
% 5 {} token are mandatory
 
  \abstract
   {We present the results of a seven-year-long radial velocity survey of a sample of 88 main-sequence and evolved stars 
   to reveal signatures of Jupiter-mass planets in the solar-age and solar-metallicity open cluster M67.
   }
   {We aim at studying the frequency of giant planets in this cluster with respect to the field stars. 
   In addition, our sample is also ideal to perform a long-term study to compare
   the chemical composition of stars with and without giant planets
in detail.
   }
   {We analyzed precise radial velocity (RV) measurements obtained 
   with the HARPS spectrograph at the European Southern Observatory (La Silla), the SOPHIE spectrograph
   at the Observatoire de Haute-Provence (France), the HRS spectrograph at the Hobby Eberly Telescope (Texas), and the HARPS-N spectrograph at 
   the Telescopio Nazionale Galileo (La Palma). 
   Additional RV data come from the CORALIE spectrograph at the Euler Swiss Telescope (La Silla).\\
   We conducted Monte Carlo simulations
   to estimate the occurrence rate of giant planets in our radial velocity survey.
   We considered orbital periods between 1.0 day and 1000 days and planet masses between 0.2\JM\, and 10.0\JM.
   We used a measure of the observational detection efficiency to determine the frequency of planets for each star. 
    }
   {All the planets previously announced in this RV campaign with their properties are summarized here: 
    3 hot Jupiters around the main-sequence stars YBP1194, YBP1514, and YBP401, 
    and 1 giant planet around the evolved star S364.
    Two additional planet candidates around the stars YBP778 and S978 are also analyzed in the present work.
    We discuss stars that exhibit large RV variability or trends individually.  
    For 2 additional stars, long-term trends are compatible with new binary candidates or substellar objects,
    which increases the total number of binary candidates detected in our campaign to 14.\\  
   Based on the Doppler-detected planets discovered in this survey, 
   we find an occurrence of giant planets of $\sim$18.0$^{+12.0}_{-8.0}$\%
   in the selected period-mass range.
   This frequency is slightly higher but consistent within the errors with the estimate for the field stars, which leads to 
   the general conclusion that open cluster and field statistics agree. 
   However, we find that the rate of hot Jupiters in the cluster ($\sim$5.7$^{+5.5}_{-3.0}$\%) is substantially higher than in the field.
   }
   {}


   \maketitle
%
%________________________________________________________________

\section{Introduction}

 In recent years, several observational campaigns have been dedicated to search for planets in clusters or stellar associations,
 where the majority of stars is considered to form.
 Stars in clusters constitute a homogeneous sample
 in age and chemical composition that is ideal for investigating the dependence of planet formation on the
 mass and the properties of the central star 
 \citep{Gonzalezhern2013, Gonzalezhern2010, Baumann2010, Johnson2010, Ramirez2010, Melendez2009}, 
 for determining the formation timescale
 and distinguishing different migration processes \citep{Dong2014,Quinn2013, Dawson2013}, 
 and finally for modeling the effects of stellar encounters on the formation 
 and evolution of planetary systems \citep{Li2015, Cai2015, Shara2014, Davies2014, Malmberg2011, Spurzem2009}.
 However, the high-precision radial velocity (RV) technique and the transit method  
 have been successful only very recently in discovering planetary-mass companions 
 around stars belonging to open clusters (OC). 
 Of the most recent results in this field,
 we mention the detection of a hot Jupiter and a multi-planetary system in the Praesepe OC \citep{Quinn2012, Malavolta2016}, 
 of a hot Jupiter in the Hyades \citep{Quinn2013},
 of two sub-Neptune planets in NGC6811 \citep{Meibom2013}, and 
 the discovery of five Jupiter-mass planets in M67 \citep{Brucalassi2014,Brucalassi2016}.
 Previous RV surveys provided detections of a long-period giant
 planet around one of the Hyades clump giants \citep{Sato2007} and a substellar-mass object in NGC2423 \citep{Lovis2007}. No significant exoplanet
 candidates have been found by either ground- or by space-based transit
 campaigns in globular clusters \citep{Gilliland2000,Nascimbeni2012}.\\
 These discoveries confirm that giant planets exist 
 in a dense cluster environment, and 
 suggest that a complete census of planet discoveries may be biased by 
 the detection limit of the instruments and the observations available today.
 Moreover, results from various simulations show that dense birth environments such as stellar clusters
 can significantly influence the planet formation process and the resulting orbital properties of the planetary systems.
 Close encounters between stars or binary companions can modify the structure of any planetary system and also subsequently generate
strong interactions between planets over very long timescales \citep{Davies2014, Malmberg2011}.
 This leads to the ejection of some planets, but it also seems to favor the conditions for the formation of hot Jupiters \citep{Shara2014}.
 Studying hot Jupiters in OCs can therefore shed light onto the long-standing problem of identifying their dynamical origin.\\
 For the past seven years we have carried out a search for massive planets around main-sequence (MS) and evolved
 stars in the OC M67.
 The scientific motivations for these studies include determining the impact of 
 a different environment on the frequency and the evolution of planetary systems with respect to field stars.
 As a long-term goal, we aim to study the connection between giant planet formation and stellar mass and chemical composition.
 M67 is the perfect target to search for planets around OC stars.
 Chemical analysis from several works \citep{Randich2006, Pace2008,Onehag2011,Onehag2014}  has shown that M67
 has a chemical composition (not only Fe, but also the other elements) that is extremely similar to solar, as close as allowed by the
 precision of the measurements. 
 In addition, the age of M67 (3-5 Gyrs), according to numerous determinations, encompasses the accepted value of the Sun,
 while the age determination for field stars is always rather uncertain.
 Finally, M67 is a rich OC, which
 gives us the opportunity to find many stellar candidates that
share similar properties and a large number of stars with
 different masses, characteristics that are essential to address the questions above.\\
 Through RV measurements 
 obtained with HARPS at La Silla-ESO, SOPHIE at OHP, HRS at HET, HARPS-N at the TNG, and CORALIE at the Euler Swiss Telescope, 
 five new giant planets have been discovered around M67 stars
 \citep{Brucalassi2014,Brucalassi2016}: three hot Jupiters around MS stars, and two long-period planets around evolved stars.
 
   In this paper, we present our RV campaign around stars of the OC M67
   considering data obtained until March 2015, and we provide
   a complete census of all the stars.\\
   The star sample is described in Sect.~\ref{sec:Sample_Obs}, RV observations and analyses are reported in Sect.~\ref{sec:RV_Obs}.
   In Sect.~\ref{sec:Results} we provide more detailed information on individual objects.
   In Sect.~\ref{sec:Planets_frequency} we report a series of simulations based on a Monte Carlo approach
   that we used to estimate the occurrence rate of giant planets in our radial velocity survey.
   Finally, we summarize our results in Sect.~\ref{sec:Conclusion}.

\section{Sample and observations}   
\label{sec:Sample_Obs}
  %_____________________________________________________________
%
\begin{table}
\caption{List of the observations: number of observed stars, total number of observations,
 number of main-sequence (MS), turn-off (TO), and giant (G) stars observed for each instrument.
 HARPS (H), SOPHIE (S), HET, CORALIE (C), and HARPS-N (HN) }             
\label{Summary_Obs}      
\centering
\resizebox{0.4\textwidth}{!}{%
\begin{tabular}{lrrrrr}     % 5 columns 
\hline\hline        
                      % To combine 4 columns into a single one 
  Instrument & H & S & HET & C & HN\\
\hline                    
 N. Stars               &  88        &  70      & 24      & 14  & 13\\%
 Observations           & 734        &  168     & 125     & 99  & 23 \\%
 MS                     & 481        &  75      & 59      &  0  & 12 \\%
 TO                     &  63        &  22      & 23      &  0  &  7 \\%
 G                      & 190        &  78      & 43      & 99  &  4\\ 

\hline                  
\end{tabular}
}
\end{table}
%
%_____________________________________________________________

 A complete description of the sample is reported in 
 our previous works \citep{Pasquini2012,Brucalassi2014}.
 We highlight here those characteristics that are
 most relevant for the aim of this paper.\\
 The original M67 sample includes a total of 88 stars with V mag between 9
 and 15, and a mass range of 0.9-1.4 \SM.\\ 
 Main-sequence stars
 have been selected following \citet{Pasquini2008}. 
 In particular, we considered those stars
 with a membership probability higher than 60\% and
 a proper motion lower than 6 mas/yr with respect to the average
 according to \citet{Yadav2008}.
 When we considered the selection of the giants, we referred
 to \citet{Sanders77} for the membership probability, 
 and the RV membership was derived
 according to \citet{Mermilliod2007} and \citet{Mathieu1986}.
 We are aware that several stars that are particularly close to the turn-off point,
 although fulfilling our selection criteria,
 have been not observed because the observation time was too short.
 
 Five different telescopes and instrument combinations have been used to obtain
 the RV measurements.
 Table~\ref{Summary_Obs} summarizes the number of the observations for each instrument.
 
 The HARPS spectrograph \citep{Mayor03} at the ESO 3.6m telescope 
 was used in high-efficiency mode (EGGS mode) 
 since the M67 stars are quite faint for this instrument.
 In this configuration the fiber has an aperture on the sky
 of 1.2 arcsec, corresponding to R=90000, and is not equipped
 with an optical scrambler. The precision is limited to 
 5-7 \ms , but the efficiency is 30-40\% better than the high-resolution mode
 (HAM mode), which allows us to obtain the highest resolving power of 115000 with a fiber aperture of 1 arcsecond.
 The spectral range covered is 378-691 nm.
 Exposure times ranged from
 10 to 45 minutes, yielding a typical signal-to-noise ratio (S/N) per resolution element of 10 to 15 for the faintest stars.
 Between January 2008 and March 2015, we gathered 734 observations of 88 stars with HARPS 
 with an associated internal precision of $\sim$ 10 \ms. 
 These observations represent the  majority of this survey, and we therefore consider HARPS as our reference.

 The SOPHIE spectrograph \citep{Bouchy06} at the OHP 1.93 m telescope was used in high-efficiency mode
 with R=40000 and a spectral range of 387-694 nm.
 We analyzed 168 SOPHIE observations of 70 M67 stars with an
 associated internal precision of $\sim$ 12 \ms.
 
 For the HRS spectrograph \citep{Tull98} at the Hobby Eberly Telescope
 we opted for a configuration
 with R=60000 and a wavelength range of 407.6-787.5 nm.
 We obtained 125 observations for 24 stars
 with a typical error bar associated with the observations of $\sim$ 25 \ms.
 During the period 2013-2015, the HET telescope was not accessible because of a telescope upgrade.
 
 Moreover, we gathered 99 RV data points for 14 giant stars
 observed between 2003 and 2005 \citep{Lovis2007}
 with the CORALIE spectrograph at the 1.2 m Euler Swiss telescope.
 
\begin{figure}[t]
 \centering
 \includegraphics[width=0.5\textwidth,angle=0]{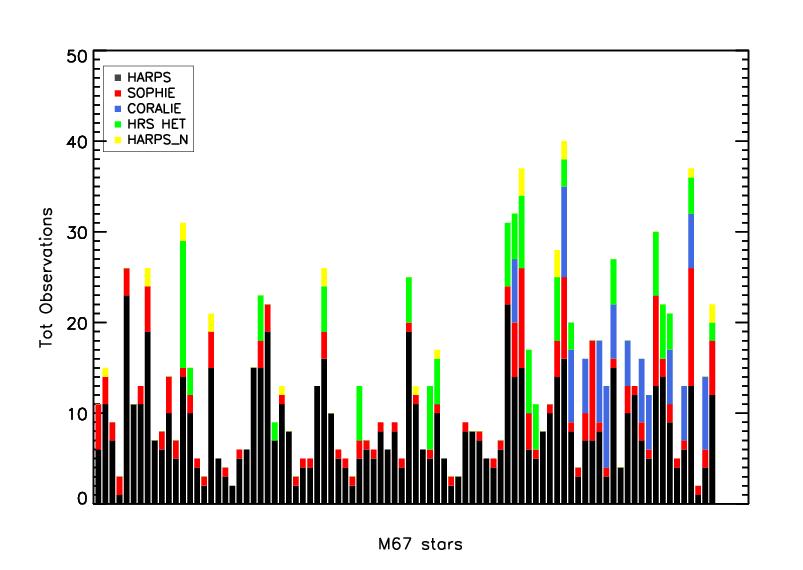}
 \caption{Histogram showing the  number of observations per star for our total sample.
 All observations from HARPS, HARPS-N, SOPHIE, CORALIE, 
 and HET HRS are included in the plot and updated to March\ 31$^{}$, 2015. 
 The order number of the stars corresponds to Table~\ref{table:observations}.}
 \label{N_Obs}
 \end{figure}

 In addition, 27 hours in service mode have been allocated 
 with the TNG on La Palma Canary Island and the HARPS-N spectrograph.
 The HARPS-N is a fiber-fed echelle spectrograph,
 similar to HARPS on the 3.6 m ESO telescope and
 covers the wavelength range between 383 to 693 nm, with a spectral resolution R=115000.
 The two HARPS fibers (object + Sky/ThAr) have an aperture on the sky of 1 arcsecond and
 are equipped with an image scrambler to provide a uniform spectrograph pupil illumination.
 Unfortunately, the observing campaign during the winter 2013-2014 was not very successful.
 The run at the TNG was impacted by bad weather and only 10 observing hours could be
 used, with 23 spectra for 13 stars.\\
 Figure~\ref{N_Obs} shows a histogram with the number of observations per star updated to March\ 31$^{}$, 2015. 
 We have obtained 13 observations per star on average:
 10 observations per star for MS stars, and 18 observations per
star for giants and turn-off stars. 
 Table~\ref{table:observations} shows the main data for the observed stars. In addition to
 the basic stellar parameters, the number of observations per star
 is given for each spectrograph and as a total. 
 
\section{RV analysis}   
\label{sec:RV_Obs}

HARPS, SOPHIE, and HARPS-N are provided with a similar automatic pipeline. 
 The spectra are extracted from the detector images and
 cross-correlated with a numerical mask. 
 For all of our stars, irrespective of the spectral type and luminosity,
 we used a G2V mask obtained from Sun spectra.
 Radial velocities are derived by fitting
 each resulting cross-correlation function (CCF) with a Gaussian
 \citep{Baranne1996,Pepe2002}.
 This real-time pipeline also provides an associated internal RV error $\sigma_{pn}$ (photon noise error).\\
 For the HRS data, the radial velocities were computed 
 using a series of dedicated routines based on IRAF\footnote{http://iraf.noao.edu/}.
 The procedures are described in more detail in \citet{Cappetta2012}.
 The different steps include the wavelength calibration
 using a Th-Ar lamp exposure acquired before and after
 each stellar spectrum, the normalization of the spectra, the
 cleaning of cosmic rays as well as telluric and sky lines, the cross-correlation 
 of the spectrum performed order by order with a G2V star template, and finally,
 the computation of the heliocentric corrections.
 The resulting CCFs were fitted 
 with a Gaussian function to determine the RVs.
 After the orders affected by telluric lines and low S/N are rejected, 
 the final RV value is given by the average value over the retained orders.
 Finally, the internal RV uncertainties are calculated by
 $\sigma_{pn}=rms(v)/\sqrt{N}$, where $v$ is the RV of the individual orders and N is the number of the orders.

\begin{figure}[t]
 \centering
 \includegraphics[width=0.45\textwidth,angle=0]{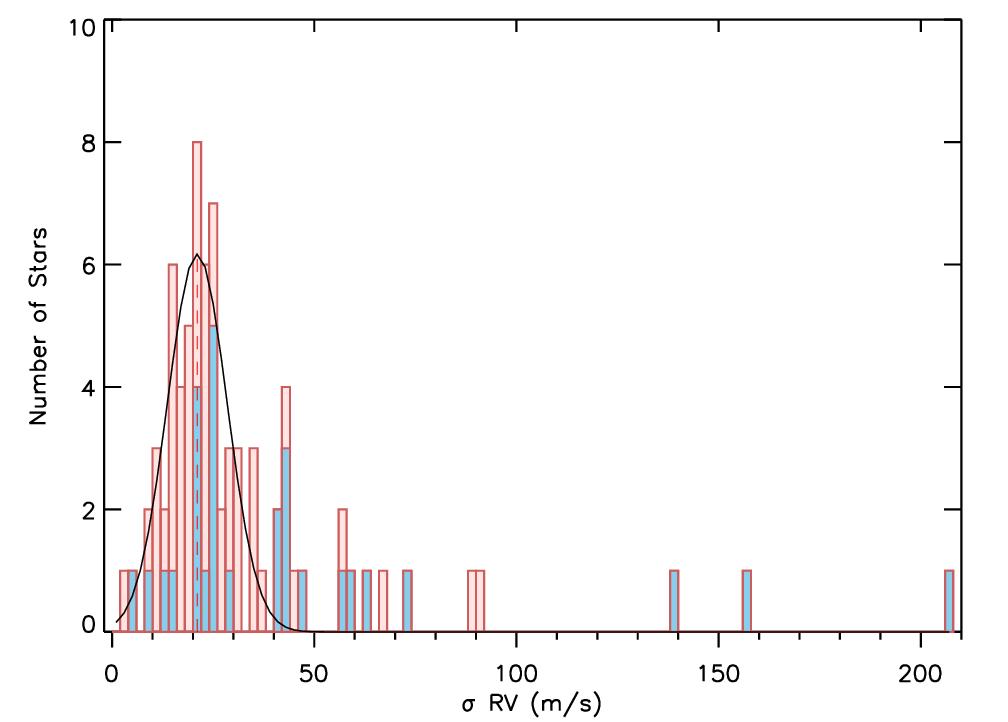}
 \includegraphics[width=0.45\textwidth,angle=0]{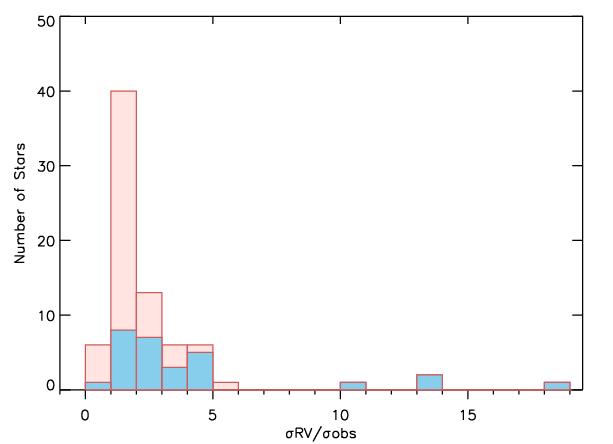}
 \caption{Top: Histogram showing the observed velocity dispersions ($\sigma_{RV}$) for the MS stars (pink) 
 and giants and turn-off stars (blue) of our sample. 
 Binary candidates and stars with high RV trend have been excluded.
 The black line represents the Gaussian fit centered at $\sim$21 \ms of width $\sigma \sim$7 \ms.
 Bottom: Histogram showing the ratio of the RV dispersion ($\sigma_{RV}$) and the average measurements error ($\sigma_{obs}$).
 The symbols are the same as in the top panel.
 }
 \label{SigmaMarch2014}
 \end{figure}

 We used nightly observations of the RV standard star HD32923
 to correct all observations for each
 star to the zero-point of HARPS \citep[as explained
 in][]{Pasquini2012} and to take any 
 instrument instability or any systematic velocity shifts between runs
(such as the modification of the SOPHIE fiber link in June 2011 \citep{Perruchot2011}
 or technical problems during the calibration phase at HARPS) into account. 
 An additional correction was applied to the SOPHIE data to
 account for the low S/N of the
 observations. For this, we corrected our
 spectra using Eq. 1 in \citet{Santerne2012}.
 Finally, we considered the total error of the RV measurements ($\sigma_{obs}$) to be 
 the sum in quadrature of all RV error sources described above.
 After all the observations for each star were corrected to the zero-point of HARPS,
 they were analyzed together.
 The combined measurement uncertainties have been compared to the observed velocity dispersions
 to evaluate the significance of any potential velocity variation and 
 to highlight any possible outliers that would suggest the presence of exoplanets.
 In Table~\ref{table:observations} the mean stellar RV of
 each star is given, together with the average associated error ($\sigma_{obs}$) 
 and the RV dispersion of the observations ($\sigma_{RV}$).
 Figure~\ref{SigmaMarch2014} shows (top panel) the histogram of the observed RV scatter ($\sigma_{RV}$) 
 for the sample stars.
 Binary candidates were excluded, and the stars with significant RV trend are not shown.
 The bulk of our observations have
 an RV scatter represented by a Gaussian distribution
 centered on $\sim21\pm7$ \ms.
 Stars with an RV variability at or above 40 \ms \ are considered very good candidates
 for low-mass companion hosts.\\
 Figure~\ref{SigmaMarch2014} also includes a second panel (bottom) 
 showing the histogram of the ratio between the observed RV scatter ($\sigma_{RV}$) 
 and the average associated error ($\sigma_{obs}$) for each star.
 Large variations for planet candidates 
 or long-term objects (see Sect.~\ref{sec:Results}) are easy to identify.
 A clear peak results for most of the stars.
 Considering only stars with $\sigma_{RV}<4\sigma_{obs}$ (the small variations) and with
 no suspected companions, we calculated a peak of $\sim$1.3 (employing a
 kernel density estimator to smooth the distribution), 
 which is a sign that we may have underestimated the measurement errors.
 However, halos of excess scatter (RV variability) are present, which are 
 most likely due to a combination of effects. In evolved
 stars some measurable intrinsic RV variability is present, 
 as shown also in \citet{Setiawan2004} and \citet{Hekker2008}, while
 for the faint MS stars the uncertainty in the measurements
 increases because of the limited S/N. Finally,
 some of the stars still have only a few observational points
 with poorly constrained RV variability, and their scatter is
 very likely the result of our low data statistics.
 \citet{Pasquini2012} also investigated whether other instrumental effects
 such as the observed flux or airmass (not included in the data
 analysis) could affect the RV measurement precision at low count levels,
 but no correlation was found. 
 In order to consider these effects during the computation of the giant planet occurrence rate (see Sect.~\ref{sec:Planets_frequency}), 
 we empirically inflated the estimated errors by adding in quadrature a term ($\sigma_{RV}'$)
 equivalent, on average, to the excess RV scatter. This is calculated by
 $\sigma_{RV}'=\sqrt{\bar{\sigma}_{RV,st}^2-\bar{\sigma}_{obs,st}^2}$,
 where $\bar{\sigma}_{RV,st}^2$ and $\bar{\sigma}_{obs,st}^2$ are
 the mean RV scatter and the mean estimated error, respectively, of stars without a trend and suspected companions.
 In Table~\ref{table:observations} the corrected associated error for each star is reported as $\sigma_{cor}$.\\  
 We studied the RV variations of our target stars by
 computing the Lomb-Scargle periodogram \citep{Scargle1982,Horne1986}. 
 This is a commonly used technique for searching for periodic sinusoidal signals
 in unevenly sampled data and allows estimates of the detection threshold to be written down for periods shorter than the duration 
 of the observations \citep{Horne1986}. 
 The significance of the sinusoid best fit of our RV values was determined by analytically calculating
 the false-alarm probability (FAP) level \citep{Horne1986}.
 Afterward, we applied Levenberg-Marquardt analysis \citep[RVLIN]{Wright2009} to fit Keplerian orbits to the radial velocity data.\\
 Radial velocity periodic variation can be caused by rotational inhomogeneities 
 related to stellar surface activity, such as plages or spots,
 including the one that is due to magnetic cycles of several years \citep{Santos2010}.
 Stellar activity can be diagnosed with spectral indicators or by monitoring the shape of the spectral lines.
 The low S/N of our observations does not provide 
 sufficient signal in the region of the sensitive Ca II H and K lines.
 We therefore followed a method similar to the one described 
 in \citet{Pasquini1991}. 
 We investigated  the presence and variability of
 chromospheric active regions  in  these  stars
 by measuring the
 variations of the core of H$\alpha$
 with respect to the continuum 
 (see Dollinger 2008 for a more
 detailed description of how the H$\alpha$ was measured).
 For each case we verified the correlation between the RVs and the bisector span 
 of the CCF \citep[calculated following][]{Queloz01} or with the FWHM of the CCF.

\section{Results}   
\label{sec:Results}

\begin{figure*}
 \centering
 \includegraphics[width=0.45\textwidth,angle=0]{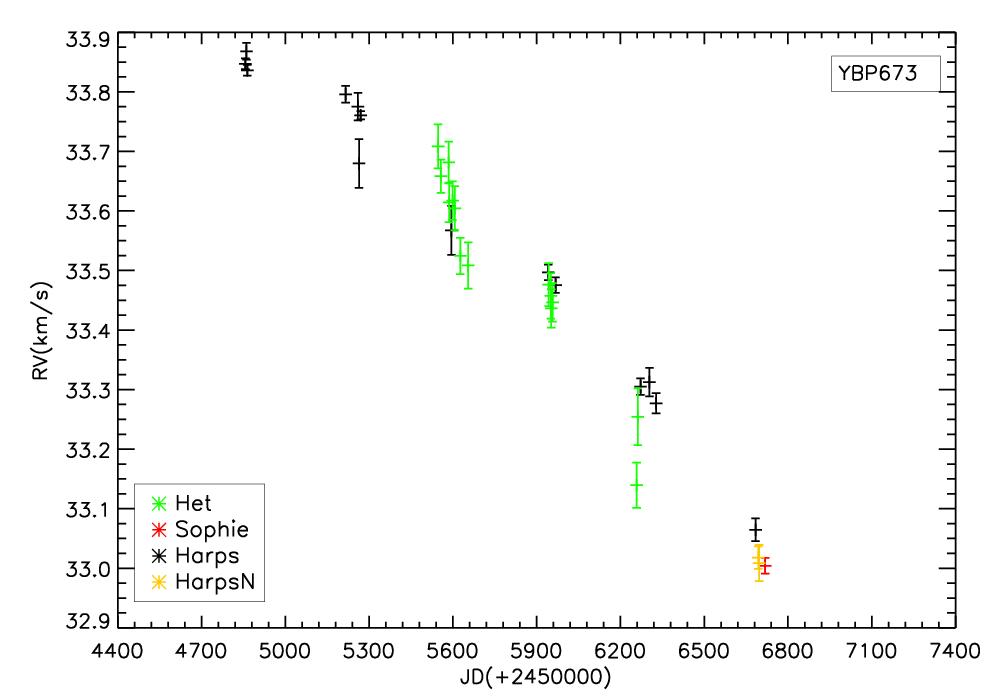}
 \includegraphics[width=0.45\textwidth,angle=0]{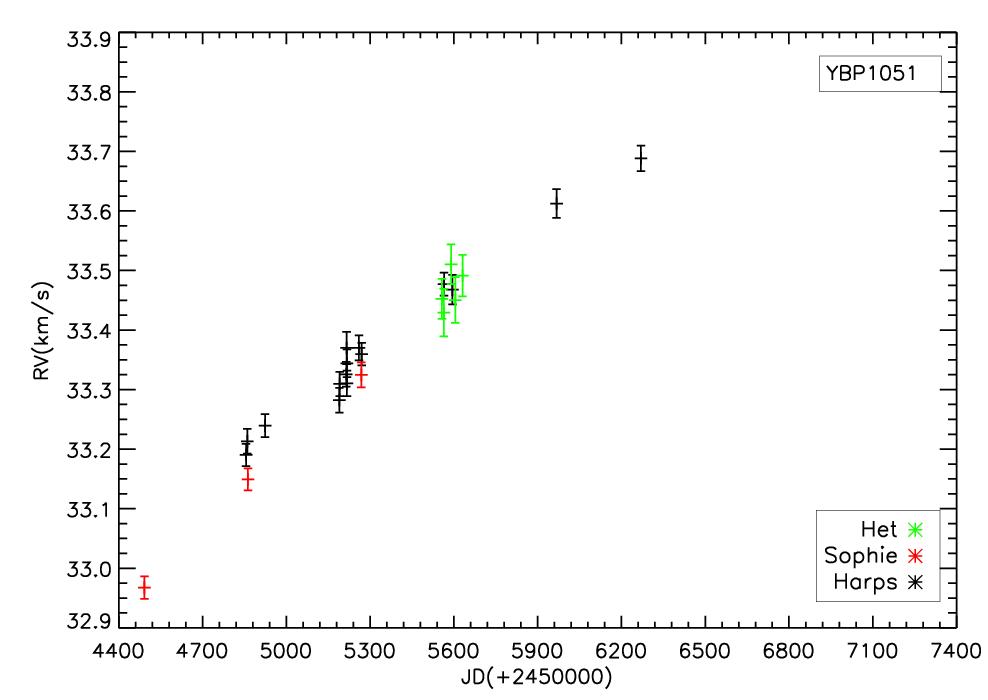}
 \caption{Stars with large RV or significant linear trends (suspected binaries).
  Black dots: HARPS measurements, red dots: SOPHIE measurements, orange dots: HARPS-N measurements,
  and green dots: HRS measurements. RV error bars represent $\sigma_{cor}$. }
 \label{NewBinaries}
 \end{figure*}

 In the following, we describe the general results of our survey 
 and discuss our analyses and interpretation 
 of the RV time series for the most relevant targets in more
detail.
 In particular, two further planet candidates are reported in this work. 
 For the other stars, we refer to Table~\ref{table:observations},
 in which the main data for the observed objects are summarized.
 Individual RV measurements for all the stars are provided as online material
 in the format listed in Table \ref{table_RVObs}.
 
 %_____________________________________________________________
%
\begin{table}
\caption{RV measurements, RV uncertainties, and instrument. 
  All RV data points are corrected to the zero-point of HARPS. 
  The errors $\sigma_{obs}$ do not include the correction for the excess RV scatter (see Sect.~\ref{sec:RV_Obs}).}             
\label{table_RVObs}      
\centering
\resizebox{0.4\textwidth}{!}{%
\begin{tabular}{llrrrl}     % 5 columns 
\hline\hline        
                      % To combine 4 columns into a single one 
  Star & BJD & RV & $\sigma_{obs}$ & Instrument\\
       & (-2450000) & (\kms) & (\kms) &  \\
\hline                    
 YBP266     &4488.509042&  33.78637     & 0.012   & SOPHIE \\%
            &4855.591058&  33.77123     & 0.012   & HARPS \\%
            &4859.605234&  33.77318     & 0.014   & HARPS \\%
            &4862.702784&  33.77678     & 0.017   & HARPS \\%
            &5189.693914&  33.79851     & 0.017   & HARPS \\

\hline                  
\end{tabular}
}
\end{table}
%
%_____________________________________________________________

\subsubsection*{\textit{Binary candidates}}

 Twelve binary candidates were previously published in \citet{Pasquini2012}.
 Here we present two other MS stars in the sample (YBP1051, YBP673) that either show significant linear trends or
 RV variations that are too large to be produced by an exoplanet
 or by a non-stellar object. The RV measurements of these stars are shown in Fig.~\ref{NewBinaries}.
 They display a peak-to-peak RV amplitude of $\sim$1.0\kms
 with an RV range spanned over more than six years.
 All binaries are highlighted in boldface in Table~\ref{table:observations}. 

\subsubsection*{\textit{Planets and planetary candidates}}
 Four planets were announced  in our previous works \citep{Brucalassi2014,Brucalassi2016}  
 for the stars YBP401, YBP1194, YBP1514, and S364.
 Table~\ref{PlanetParam} reports the main stellar characteristics and the 
 resulting updated orbital parameters of the Keplerian fit for all these stars.
 Another two stars (S978 and YBP778) show significant indication for the presence of Jovian-mass companions.
  
\begin{figure}
  \vspace{-5pt}  
 \centering
  \hspace{-16pt}
 \includegraphics[width=0.48\textwidth,angle=0]{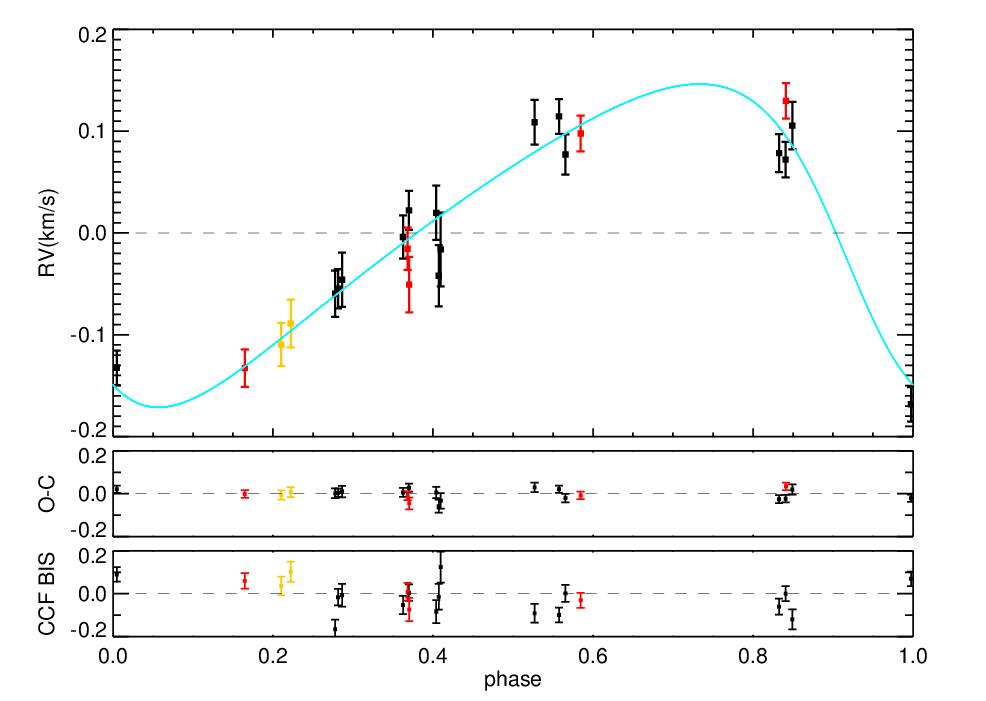}
\vspace{-5pt} 
 \caption{
  Phased RV measurements and Keplerian best fit, best-fit residuals, and
  bisector variation for YBP778. Same symbols as in Fig.~\ref{NewBinaries}.} 
 \label{Fit_YBP778}
 \end{figure}

 \textbf{YBP778}. Twenty-one RV measurements have been obtained for the MS star YBP778 since 2009: 
 15 with HARPS, the others with SOPHIE and HARPS-N.
 The typical S/N is 10 and the average measurement uncertainty is
 $\sim$17\ms\, for HARPS and HARPS-N, and $\sim$10\ms\, for SOPHIE.
 A clear periodic signal can be seen in the periodogram at $\sim$398 days (see Fig.~\ref{Periodograms}).
 Thus a single-planet Keplerian model was adjusted to the data, 
 and the best-fit solution corresponds to a signal with a period of
 410.4$\pm$6.2 days, a semi-amplitude 158.8$\pm$21.4\ms\, (see Fig.~\ref{Fit_YBP778}), and an eccentricity of 0.27$\pm$0.11.
 Although no peak is present in the periodograms of the activity index or the bisector span, nor in the CCF FWHM (see Fig.~\ref{Periodograms}),
 we detected an anticorrelation between the RVs and the bisector span with a Pearson correlation coefficient of -0.56,
 and a slight correlation between the radial velocities and the CCF FWHM (see Fig.~\ref{BIS_FWHM_Ha}).
 The correlation disappears when we consider the RV residual.\\  
 We therefore cannot completely exclude stellar activity (magnetic cycles)
 or a binary companion with low $\sin(i)$ as the cause of the RV variation at this point, although
 in the recent study of \citet{Geller2015}, YBP778 is also classified as a single member.
 Interesting, \citet{Pace2012b} noted that this star is significantly underabundant in lithium compared to
 the lithium abundances as a function of stellar mass for MS stars in M67.
 This could argue in favor of the possibility that the star might indeed host a planet \citep[see, e.g.,][]{Deal2015}.
 However, \citet{Pace2012b} warned that the distance of this star to the isochrone in the color-magnitude diagram (CMD; $\sim$0.06 mag)
 could also suggest the presence of a stellar or substellar companion.
 
\begin{figure}
 \vspace{-5pt}
 \centering
 \hspace{-16pt}
 \includegraphics[width=0.48\textwidth,angle=0]{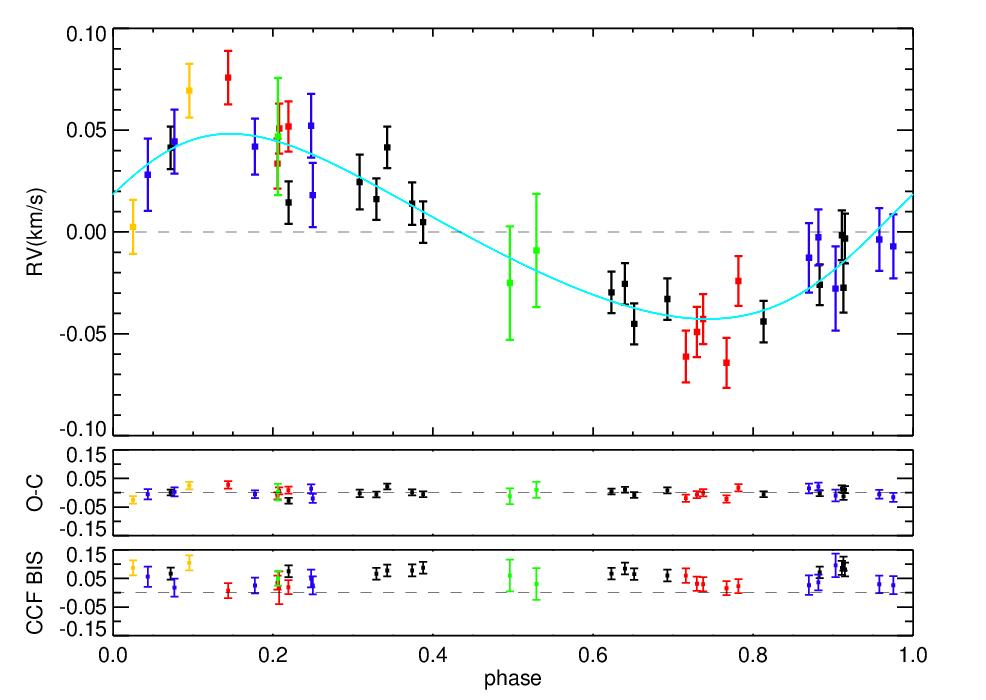}
 \vspace{-5pt}
 \caption{
  Phased RV measurements and Keplerian best fit, best-fit residuals, and
  bisector variation for S978. Same symbols as in Fig.~\ref{Fit_YBP778}. Green dots: HRS measurements,
  blue dots: CORALIE measurements.}
 \label{Fit_S978}
 \end{figure}

 \textbf{S978}. The star is a K4 red giant and was observed 40 times over more than five years with HARPS, HRS, SOPHIE, and HARPS-N.
 The average RV uncertainty is $\sim$3.0\ms\, for HARPS and SOPHIE,
 $\sim$26.0\ms\, for HRS, and $\sim$8.0\ms\, for HARPS-N . Ten additional RV measurements
 were obtained with CORALIE between 2003 and 2005, with a
 mean measurement uncertainty of $\sim$12.0\ms.
 The strongest peak in the periodogram of the RV time series lies at about 510.4 days (see Fig.~\ref{Periodograms}).
 Aliases are also present in the periodogram, but at half of the power of the main signal.
 The bisector inverse slope as well as the activity indicator do not present any significant variation at the period of 510.4 days.
 We fitted a single-planet Keplerian orbit to this signal (see Fig.~\ref{Fit_S978}) and found an orbital
 solution whose parameters are reported in Table~\ref{PlanetParam}.
 The dispersion of the residuals is $\sigma$(O-C)= 12.90\ms\, 
 and the periodogram of the residuals reveals some structures, but no significant peaks (see Fig.~\ref{Periodograms}).
 
 Several studies indicate that M67 stars in general have a low level of chromospheric activity. \citet{Pace2004} computed activity levels for M67 and other clusters. 
 The paper shows that, for instance, the young  Hyades and Presaepe stars on average have a chromospheric emission flux in the CaII K line of $\mathrm{<F^{'}_{K}> \sim 2.1 \cdot 10^{6}}$ $\mathrm{erg\, cm^{-2}\, s^{-1}}$ and
 $\mathrm{<F^{'}_{K}> \sim 2.43 \cdot 10^{6}}$ $\mathrm{erg\, cm^{-2}\, s^{-1}}$, respectively, while M67 stars have
 $\mathrm{<F^{'}_{K}> \sim 0.5 \cdot 10^{6}}$ $\mathrm{erg\, cm^{-2}\, s^{-1}}$, which is not enough to explain 
 the high RV variations we observe.
 In a survey of the Ca II H and K core strengths of a sample of 60 solar-type stars in M67, 
 \citet{Giampapa2006} found that the distribution of the HK index (a measure of the strength of the chromospheric
 H and K cores) is broader than the distribution seen in the contemporary solar cycle. 
 Significant overlap between the HK distribution of the solar cycle and that for the Sun-like stars
 in M67 is seen with over 70\% of the solar analogs exhibiting
 Ca II H+K strengths within the range of the modern solar cycle.
 About $\sim$10\% are characterized by high activity in excess of the solar maximum values, 
 while approximately 17\% have values of the HK index lower than the solar minimum.
 Of these, none of the stars showing enhanced activity is present in our final sample.
 In a following work, the same authors reported 
 the results of the analysis of high-resolution photospheric line spectra obtained with the UVES
 instrument on the VLT for a subset of 15 solar-type stars in M67 selected by \citet{Giampapa2006}.
 They found upper limits to the projected
 rotation velocities that are consistent with solar-like rotation (i.e., vsini $\leq$ 2-3 \kms) 
 for objects with Ca II chromospheric activity within the range of the contemporary solar cycle. 
 Two solar-type stars in their sample exhibit chromospheric emission well in excess of even the solar maximum values: Sand747 and Sand1452.
 In one case, Sand747, the authors found it to be a spectroscopic binary. 
 The other star, Sand1452, was also present in our original sample, 
 but we discovered that the object was also a binary system \citep{Pasquini2012}. 
 Furthermore, of the 15 solar-type stars analyzed in \citet{Reiners2009}, 
 seven stars are in common with our sample (in particular our planet-hosts YBP1194 and YBP1514).
 These objects are slow rotators (vsini $\leq$ 2 \kms) and have
Sun-like HK values.\\
 \citet{Melo2001} used FEROS spectrograph observations to determine accurate projected rotational velocities vsini 
 for a sample of 28 MS, turn-off, and giant stars belonging to M67.
 They found that the stars show similar values of vsini depending on their position
 in the CMD. Early MS G stars have a rotational velocity twice higher
 than the Sun, and they show a possible trend with (B-V) color, with redder colors corresponding to lower vsini.
 The stars at the turn-off are the fastest rotators, with vsini between 6.3 and 7.6 \kms, while stars just above
 the turn-off are already significantly slower, with values between 4.6 and 4.9 \kms.
 Along the subgiant branch rotation tends to drop down, and for stars with (B-V)$>$1  only upper
 limits can be found, including for the clump stars (vsini $\leq$1.5 \kms). 
 Most of the stars in this last group are in common with our sample.\\
 To rule out activity-related rotational modulation as the cause of the RV variations in our data, 
 we investigated chromospheric activity by measuring the variations of the core of H$\alpha$ 
 with respect to the continuum. 
 Using public data with known correlations between activity indicators and RVs, 
 we verified that such correlations could be still recorded at low S/N level of the M67 spectra.    
 Of our targets, S364 and S978 show a variability in H$\alpha$ of
 2\%, YBP1514 and YBP1194 of 3\%  without significant periodicity and YBP401 of 4\%,
 exhibiting all a very low activity level, while for YBP778 H$\alpha$ variability 
 is slightly higher, with a value of 7\%.
 The fact that these stars are of solar age and that 
 our research is focused on finding giant planets with an expected RV variability of tens of \ms  
              makes the contamination by activity negligible. 

\subsubsection*{\textit{Stars with long-term RV variability}}
 \begin{figure*}
 \centering
 \includegraphics[width=0.4\textwidth,angle=0]{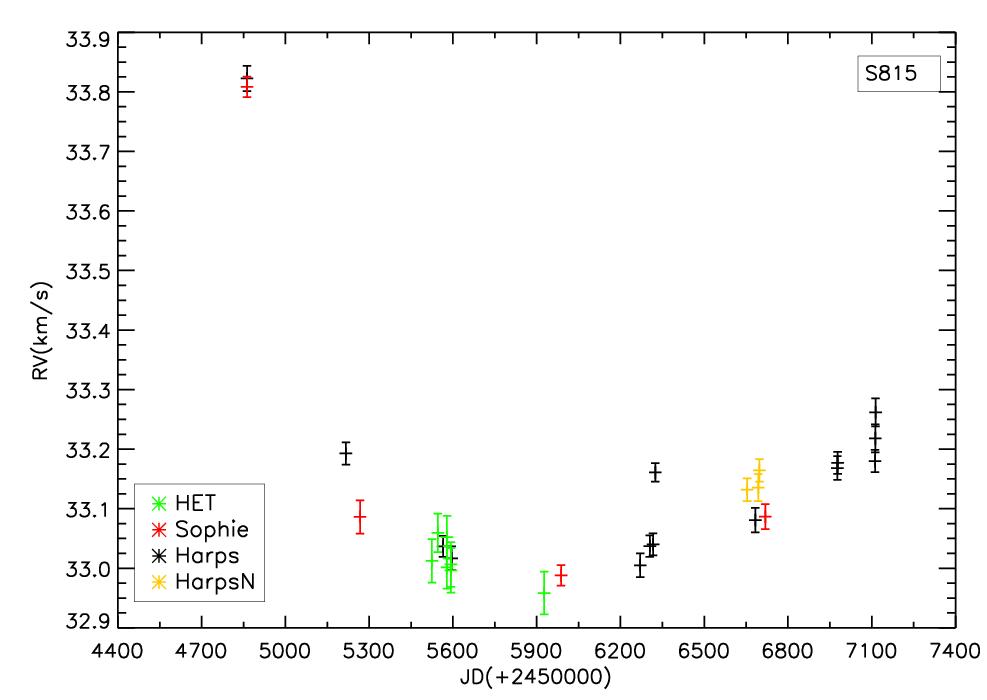}
 \includegraphics[width=0.4\textwidth,angle=0]{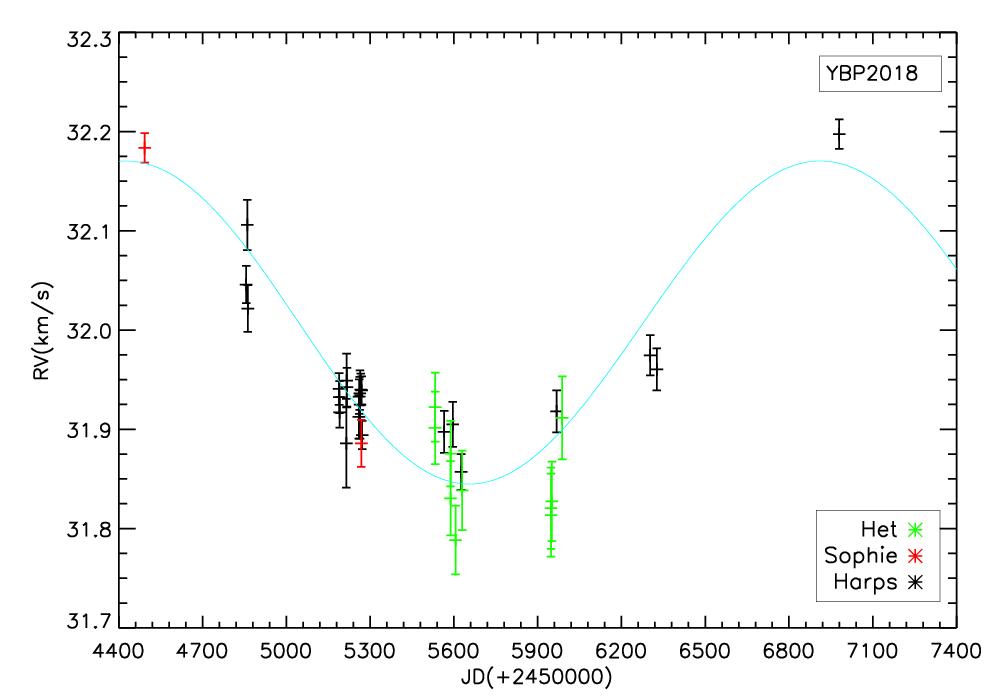}
 \caption{RV time series for S815 (left) and YBP2018 (right).
  For this latter a tentative Keplerian solution is overplotted in blue.
  Same symbols as in Fig.~\ref{NewBinaries}.}
 \label{NewBinaries_2}
 \end{figure*}
\begin{figure*}
 \centering
 \includegraphics[width=0.4\textwidth,angle=0]{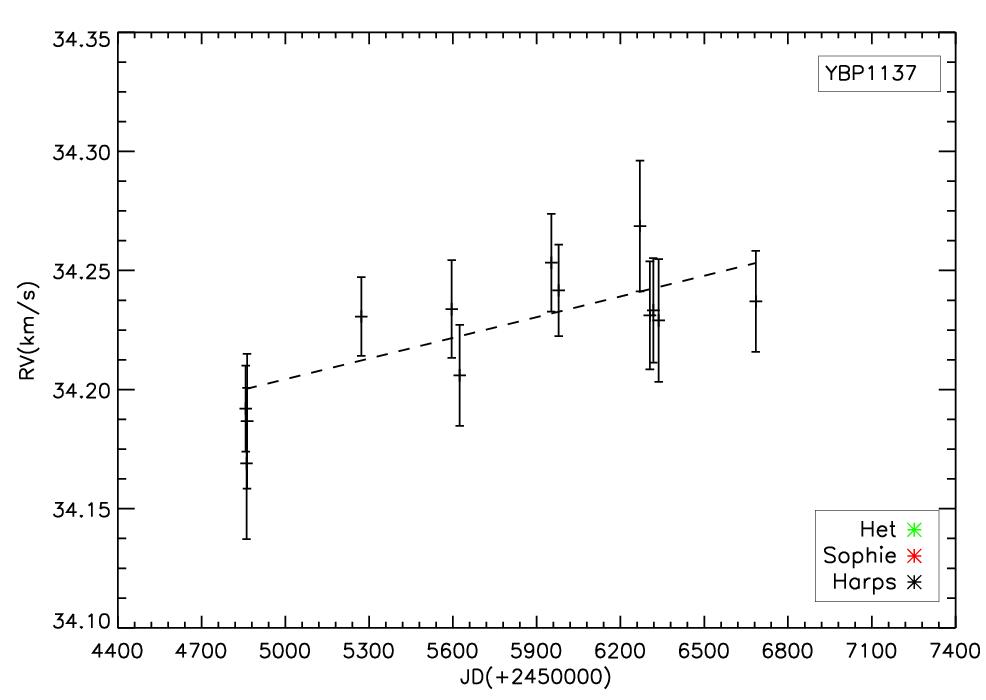}
 \includegraphics[width=0.4\textwidth,angle=0]{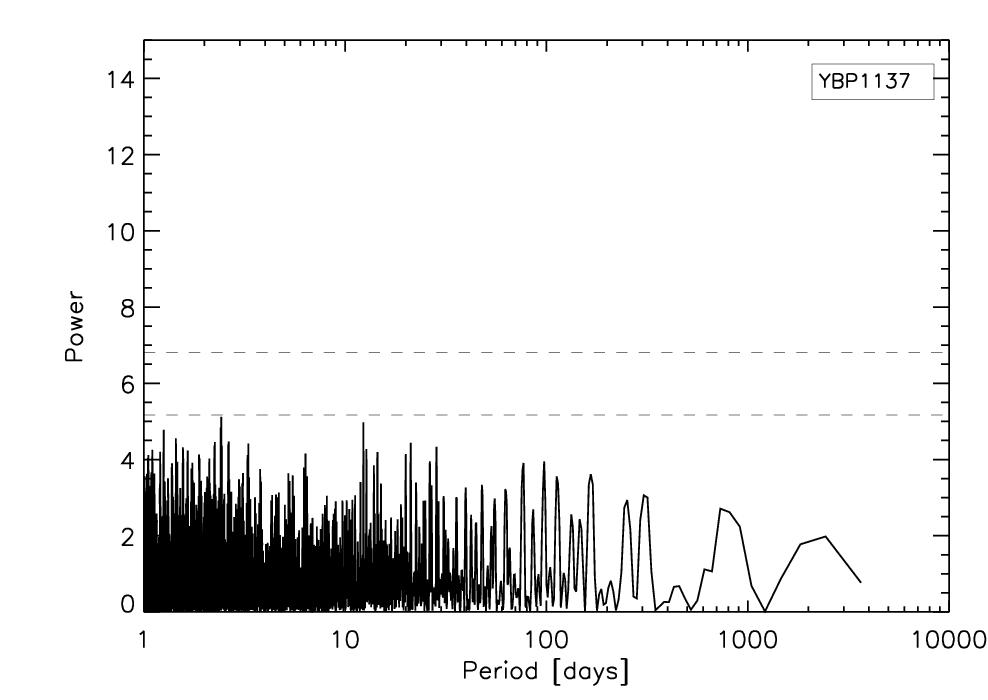}
 \caption{Left: RV time series with the fitted linear trend of YBP1137 overlaid.
          Right: Periodogram of the residuals after the RV trend has been removed.
          Same symbols as in Fig~\ref{NewBinaries}.}
 \label{Trendstars_2}
 \end{figure*}
\begin{figure*}
 \centering
 \includegraphics[width=0.3\textwidth,angle=0]{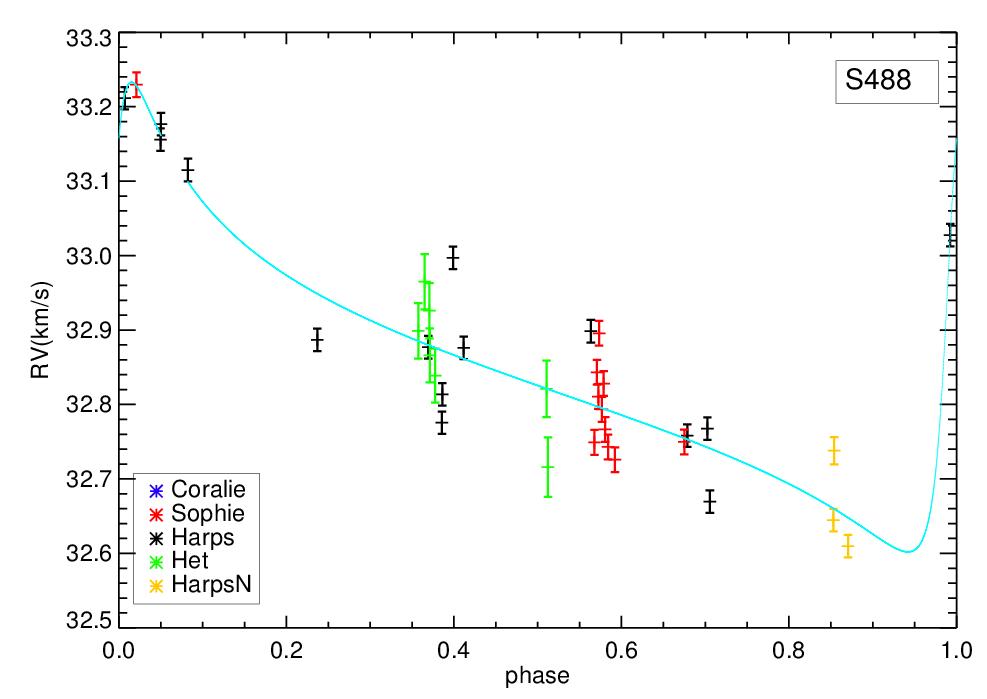}
 \includegraphics[width=0.3\textwidth,angle=0]{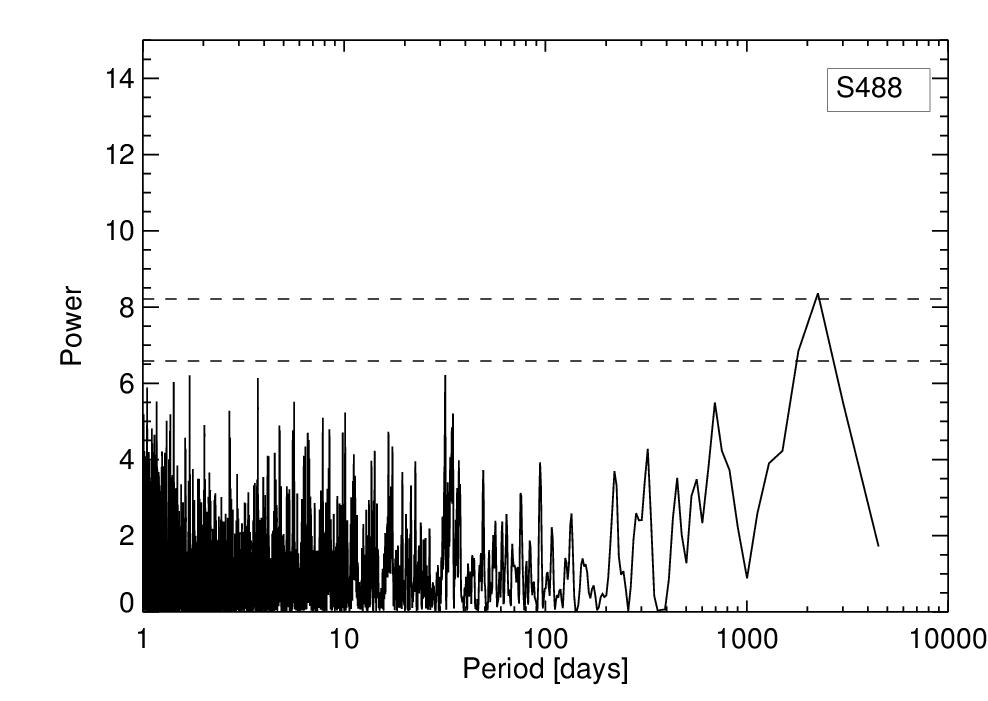}
 \includegraphics[width=0.3\textwidth,angle=0]{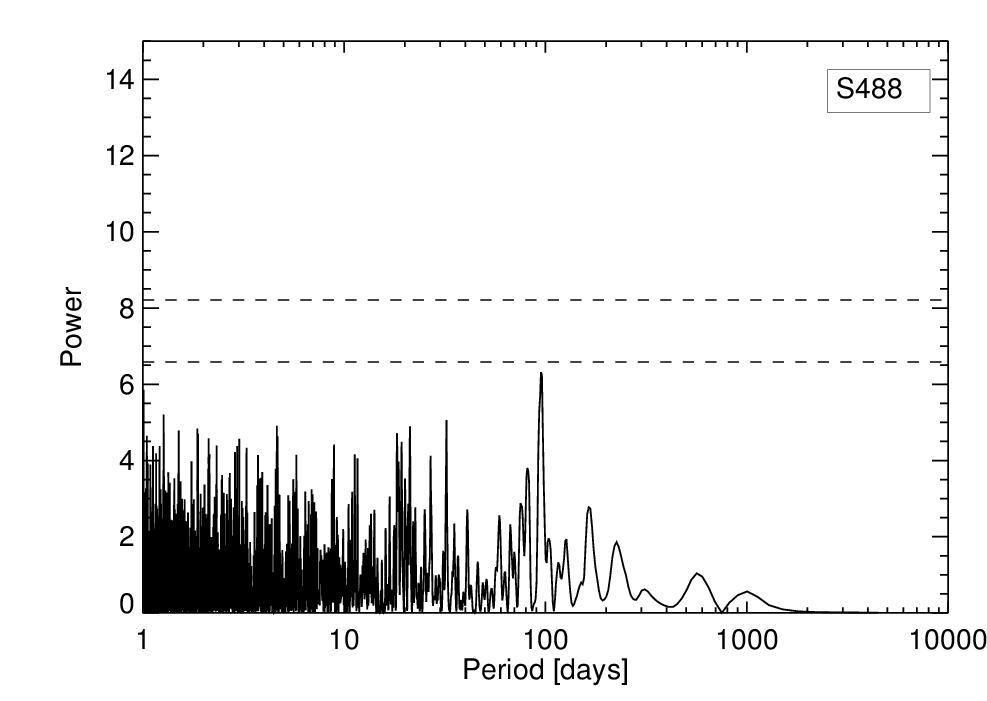}
 \caption{Phased RV measurements and Keplerian best fit (left), periodogram of the RV data (center),
 and periodogram of the residuals for S488.}
 \label{S488}
 \end{figure*}

 The turn-off star S815 shows a peak-to-peak 
 RV variation of the order of $\sim$700 \ms (see Fig.~\ref{NewBinaries_2}).
 This star is retained in the single-star sample, 
 although the amplitude of RV values is possibly too high for a planet.
 Moreover, the RV measurements would indicate a rather high eccentricity.\\
 The MS star YBP2018 presents peak-to-peak
 RV variations on the order of $\sim$400 \ms and appears to have almost completed one orbit.
 Figure~\ref{NewBinaries_2} shows the RV measurements with a tentative Keplerian solution, resulting in an orbiting object with
 a minimum mass of $\sim$11.0\Mj \, and a period of $\sim$2487 days.
 The residuals have an rms of $\sim$30.0\ms, but the periodogram of the residuals does not show any clear periodicity.
 Extensive follow-up over several years is required for this star to understand the nature of the companion candidate.\\
 \begin{figure*}
 \centering
  \includegraphics[width=0.4\textwidth,angle=0]{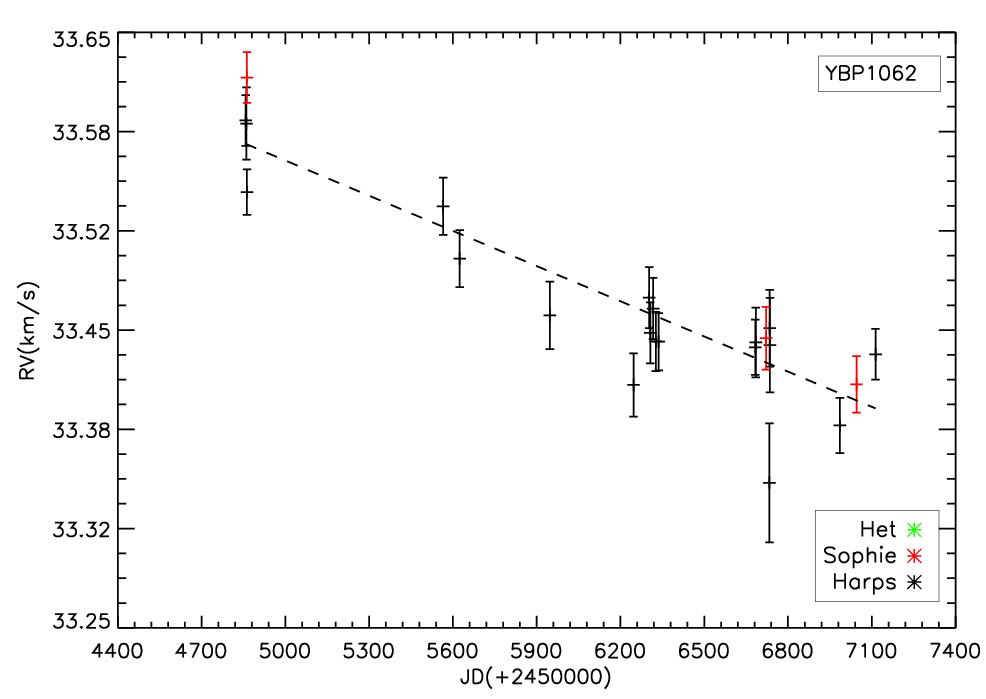}
 \includegraphics[width=0.4\textwidth,angle=0]{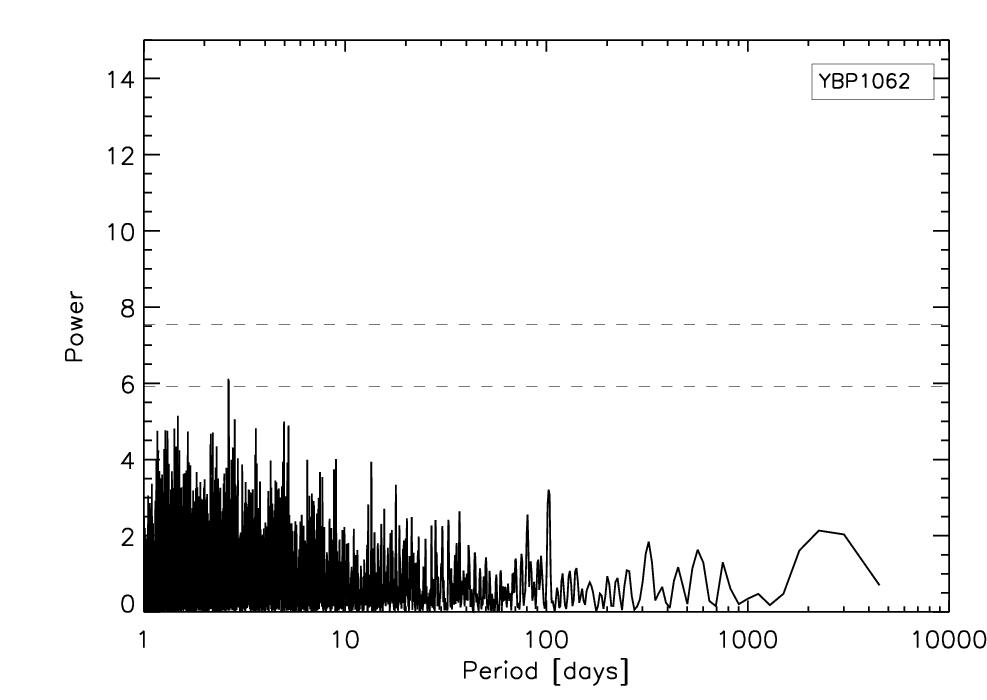}
 \caption{Left: RV time series with the fitted linear trend of YBP1062 overlaid.
          Right: Periodogram of the residuals after the RV trend has been removed.
          Same symbols as in Fig~\ref{NewBinaries}.}
 \label{Trendstars_1}
 \end{figure*}
 The MS stars YBP1062 and YBP1137 exhibit a trend in RV measurements of -28.75\ms/yr and 10.57\ms/yr.
 Figures~\ref{Trendstars_1} and ~\ref{Trendstars_2} show the RV time series with the fitted linear trend overlaid.
 Although the residuals of the linear fit present an RV variability of 33.36\ms for YBP1062 and 18.00\ms for YBP1137,
 the periodogram of the residuals for the two stars does not reveal any significant peaks.\\
 The evolved star S488 also shows a high RV variability when compared to the measurement errors but 
 caution is necessary with this star, which is located at the tip on the red giant branch.
 \citet{Dupree1999} found a strong Ca II emission-line
 with no visible change in the line asymmetries with time
 for this star, suggesting the presence of outward mass motions.\\
 When we compute a Lomb-Scargle periodogram, a peak is present at $\sim$2257 days.
 At this period, the measurements baseline is not yet long enough to constrain any significant orbital solution.
 A tentative Keplerian curve is given in Fig.~\ref{S488} and
corresponds to an object with a minimum mass 
 of $\sim$14.0\Mj \, and a period of $\sim$2332 days.
 The residuals have an rms amplitude of $\sim$55\ms , and when the main signal is removed, the periodogram of the
 residual shows a peak at $\sim$95.03 days with
 a lower significance close to a 0.05 FAP level. 
 Additional observations are needed to draw further conclusions.

 \begin{table*}
\caption{Top: Stellar parameters of the M67 stars that are newly found to host planet candidates.
Bottom: Orbital parameters of the planetary companions. \ensuremath{P}: period, \ensuremath{T}: time at periastron passage,
 \ensuremath{e}: eccentricity, $\omega$: argument of periastron, \ensuremath{K}: semi-amplitude of the RV curve, 
 \ensuremath{m\sin{i}}: planetary minimum mass,
 $\gamma$: average radial velocity, and $\sigma$(O-C): dispersion of the Keplerian fit residuals.  }
\label{PlanetParam}
\centering
\resizebox{1.0\textwidth}{!}{%
\vspace{-5pt}
\begin{tabular}{lrrrrr}
%\hline\hline
\hline
\textbf{Parameters} &YBP401&YBP1194&YBP1514&SAND364&SAND978 \\
\hline
$\mathrm{\alpha}$ $(\mathrm{J2000})$&  08:51:19.05 &   08:51:00.81   &  08:51:00.77 &  08:49:56.82 &  08:51:17.48 \\
$\mathrm{\delta}$ $(\mathrm{J2000})$& +11:40:15.80 &  +11:48:52.76   & +11:53:11.51 & +11:41:33.00 & +11:45:22.69  \\
Spec.type       &F9V    &G5V& G5V& K3III& K4III\\
\ensuremath{m_{\mathrm{V}}}  $[\mathrm{mag}]$          &13.70\tablefootmark{a}            &14.6\tablefootmark{a}           &14.77\tablefootmark{a}         &9.8\tablefootmark{b}           &9.71\tablefootmark{b} \\
\ensuremath{B-V}             $[\mathrm{mag}]$          &0.607\tablefootmark{a}            &0.626\tablefootmark{a}          &0.680\tablefootmark{a}         &1.360\tablefootmark{b}         &1.370\tablefootmark{b}   \\
\ensuremath{M\star}           [\ensuremath{M_{\odot}}] &1.14$\pm$0.02\tablefootmark{d}    &1.01$\pm$0.02\tablefootmark{d}  &0.96$\pm$0.01\tablefootmark{d} &1.35$\pm$0.05\tablefootmark{d} &1.37$\pm$0.02\tablefootmark{d} \\
%%\ensuremath{R\star}          [\ensuremath{R_{\odot}}] &0.99$\pm$0.02\tablefootmark{d}    &0.89$\pm$0.02\tablefootmark{d}  &21.8$\pm$0.7\tablefootmark{d}\\
$\logg$                      $[\mathrm{cgs}]$          &4.30$\pm$0.035\tablefootmark{f}   &4.44$\pm$0.05\tablefootmark{e}  &4.57$\pm$0.06\tablefootmark{g} &2.20$\pm$0.06\tablefootmark{h}  &1.80$\pm$0.09\tablefootmark{i} \\
$\Teff$                      $[\mathrm{K}]$            &6165$\pm$64\tablefootmark{f}      &5780$\pm$27\tablefootmark{e}    &5725$\pm$45\tablefootmark{g}   &4284$\pm$9\tablefootmark{h}    &4200$\pm$21\tablefootmark{i}\\
%%$\moh$                      $[\mathrm{dex}]$          &0.023$\pm$0.015\tablefootmark{d}  &0.02 $\pm$0.05\tablefootmark{d} &$-$0.02$\pm$0.04\tablefootmark{e}\\
\hline
\ensuremath{P}        $[\mathrm{days}]$& 4.087$\pm$0.003      & 6.960$\pm$0.001     & 5.118$\pm$0.001     & 120.951$\pm$0.453       & 511.21$\pm$2.04       \\
\ensuremath{T}        $[\mathrm{JD}]$  & 2456072.4$\pm$0.6   & 2455679.9$\pm$0.4 & 2455986.3$\pm$0.3   & 2456231.22$\pm$4.26     & 2456135.92$\pm$21.23     \\
\ensuremath{e}                         & 0.16$\pm$0.08        & 0.31$\pm$0.08       & 0.27$\pm$0.09       & 0.35$\pm$0.10           & 0.16$\pm$0.07          \\ 
$\omega$              $[\mathrm{deg}]$ & 343.33$\pm$62.12     & 109.16$\pm$20.16    & 328.58$\pm$17.78   & 254.62$\pm$ 15.91       & 291.68$\pm$35.85       \\
\ensuremath{K}        [\ms]            & 49.29$\pm$5.50       & 37.35$\pm$4.55      & 50.47 $\pm$3.90     & 56.94$\pm$4.26          & 45.48$\pm$3.65          \\
\ensuremath{m\sin{i}} [\Mj]            & 0.42$\pm$0.05        & 0.33$\pm$0.03       & 0.40$\pm$0.35       & 1.57$\pm$0.11           &  2.18$\pm$0.17          \\
$\gamma$              [\kms]           & 33.179$\pm$0.004     & 34.184$\pm$0.003    & 34.058$\pm$0.004    & 33.188$\pm$0.019        & 34.567$\pm$0.008        \\
$\chi_{red}^2$                         & 0.97                 & 0.95                 & 0.93                & 1.08                    &  1.57                   \\
$\sigma$(O-C)         [\ms]            & 12.31                & 11.51               & 14.24               & 15.93                  & 12.90                  \\
\hline
\end{tabular}
}
\tablefoot{\tablefoottext{a}{\citet{Yadav2008}.}
\tablefoottext{b}{\citet{Montgomery1993}.}
\tablefoottext{c}{\citep{Sanders77}.}
\tablefoottext{d}{\citet{Pietrinferni2004} and \citet{Girardi2000}.}
\tablefoottext{e}{\citet{Onehag2011}.}
\tablefoottext{f}{\citet{Pasquini2008} and \citet{Pace2012}.}
\tablefoottext{g}{\citet{Smolinski2011} and \citet{Lee2008}.}
\tablefoottext{h}{\citet{Wu2011}.}
\tablefoottext{i}{\citet{Jacobson2011}.}
}

\end{table*}

\section{Planet frequency}
\label{sec:Planets_frequency}

 A series of simulations based on a Monte Carlo approach have been carried out to
 determine detection limits for the RV data as a function of planet mass and planet period,
 and to derive a trustworthy estimate of the occurrence rate of giant planets for our RV survey. 
 Understanding the frequency of different types of planets around stars of different
 mass can provide important clues about the processes of planet
 formation and evolution. Moreover, the direct comparison with similar analyses on field star samples
 can reveal important indications about the influence of the
stellar birth environment on the evolution of planetary systems.\\
 In general, the determination of the detection efficiency for an RV survey is based on the ability to recover
a planetary signal with a given level of statistical significance.
 This translates into lower limits on the detectable companion mass as a function of orbital period.
 Several authors \citep[e.g.,][]{Cumming2008, Howard11} 
 have presented statistical analyses of their planet surveys
and discussed different methods to derive
 limits on companion detectability and unbiased distributions for a number of RV planet surveys.   
 They used mainly two approaches: one based on $\chi^{2}$ and F-tests to detect 
 excess residuals above an assumed Gaussian noise \citep{Lagrange2009b,Sozzetti2009},
and the other approach is based on a periodogram analysis 
 to identify significant periodicity \citep{Cumming2004,Narayan2005,Mortier2012}.
 For our project, the second approach was selected for all the stars 
 with enough data points for a reliable periodogram analysis.
 For the stars where the periodogram was not feasible because
of the combined effect of poor sampling and the small number of observations per star,
 we evaluated the detection limit using the variability of the RV values with respect to the measurement errors.

 \subsection{Method}
 \label{sec:Methodology}
 
 In order to derive the detectability of planetary signals, we computed synthetic datasets,
 simulating a series of mass and period data pairs ($\sim$10$^{6}$ values) 
 of our 'potential planets',
 which are uniformly distributed in a logarithmical mass range of 0.2-10.0 \Mj\, and 
 in a linear period range of 1.0-1000 days. 
 From the real measurements of each star we retain the observation dates (expressed in Baricentric Julian Date) 
 and the corresponding measurement errors, which we inflated as explained in Sect.~\ref{sec:RV_Obs}.
 For random choices of
 the planet mass-period pairs and assuming eccentric orbits, we calculated the contribution to the radial velocity amplitude K
 from a giant planet using the relation
 \begin{equation}
 \label{Kcirc}
 K= \left( \frac{2\pi G}{P} \right)^{1/3} \left( \frac{m_psin(i)}{\Mj}\right) \left(\frac{M_{\star}}{\SM}\right)^{-2/3}
 \left(\frac{1}{\sqrt{1-e^{2}}}\right)
 ,\end{equation}
 \
 \par\noindent
 where $M_{\star}$ is the mass of the host star that we obtained from isochrone fitting.
 A random distribution of the orbit inclination was also assumed, and
 the eccentricity $e$ was allowed to vary randomly between $0.0<e<0.5$.  
 Then, we derived the synthetic planetary signals following
 the equation for eccentric Keplerian orbits,
 \begin{equation}
 \label{RVs}
 %V_{r}(t)= Ksin\left(\frac{2\pi}{P}(t-t_{0})\right) + c
 V_{r}(t)= acos\nu(t)+bsin\nu(t)+ c
 ,\end{equation}
 \
 \par\noindent
 where $a=Kcos\omega$, $b=-Ksin\omega$ and $c=Kecos\omega + \gamma$.
 Here, $K$ is the RV amplitude, $e$ the eccentricity, $\omega$ 
 the longitude of periastron, $\gamma$ the systemic velocity of the system,
 $\nu(t)$ the true anomaly, which is a function of t, P, and
e, and $t_{0}$
 is the time of periastron passage.
 The latter was selected randomly in the time span of the observations, and $\gamma$ 
 was allowed to vary around the RV value of the M67 cluster obtained in \citet{Pasquini2012}.
 We then degraded the obtained RVs by adding noise corresponding to the actual 
 measurement error of the real observations.
 
 \subsubsection{Stars with at least ten data points}
 
 A periodogram-based analysis \citep{Scargle1982} was applied to each synthetic dataset
 to verify whether the simulated planet was observable.
 This approach was chosen for all the stars with enough measurements for a reliable periodogram analysis.
 We note that this is the same procedure as we applied for the analysis of the RV measurements in Sect.~\ref{sec:RV_Obs}.\\
 For a given mass and period values, we considered a planet to
be detected 
 if we could obtain a signal from the periodogram analysis that
had a power higher than the power associated with a 0.01 FAP.
 For each star 10$^{6}$ trials were used.\\
 In the two-dimensional space of orbital period and planet mass,
 we consider a grid of period and log-spaced mass cells 
 within which the planet detectability (or detection efficiency, $E_{ij}$) is individually computed as 
 the number of the potentially detected planets with respect to the total simulated planets in each bin of mass and period. 
 On the base of this analysis,
 the detection efficiency was carried out for 14 MS stars and 10 evolved stars.
 Figures~\ref{PlEffMS} and \ref{PlEffGTO} show the detection efficiencies for the 24 stars
 as color contours from 0.0 to 1.0. 
 The red regions have a 100\% detectability of planets, and blue regions have low detectability.
 This means that in red regions (high detectability) 
 all potential planets at the given period and mass would be detected.
 In the blue regions (low detectability) 
 the detection efficiency decreases, and we are not able to discover 
 their presence at high confidence.
 In Figs.~\ref{PlEffMS} and~\ref{PlEffGTO}, 
 the resulting distribution of the planet detection efficiency
clearly reduced with decreasing planet mass and with increasing period. This is due to the insufficient number of measurements,
 to the distribution of the observations, and to the weak RV signal.
 Moreover, windows of poor detectability in the period-mass grid occur for two main reasons: the data structure (number of observations and phase coverage),
 and the one-year seasonal period that affects any ground-based observing program.\\  
 
 \subsubsection{Stars with poor sampling and few observations}
 
\begin{figure}[t]
 \centering
 \includegraphics[width=0.4\textwidth,angle=0]{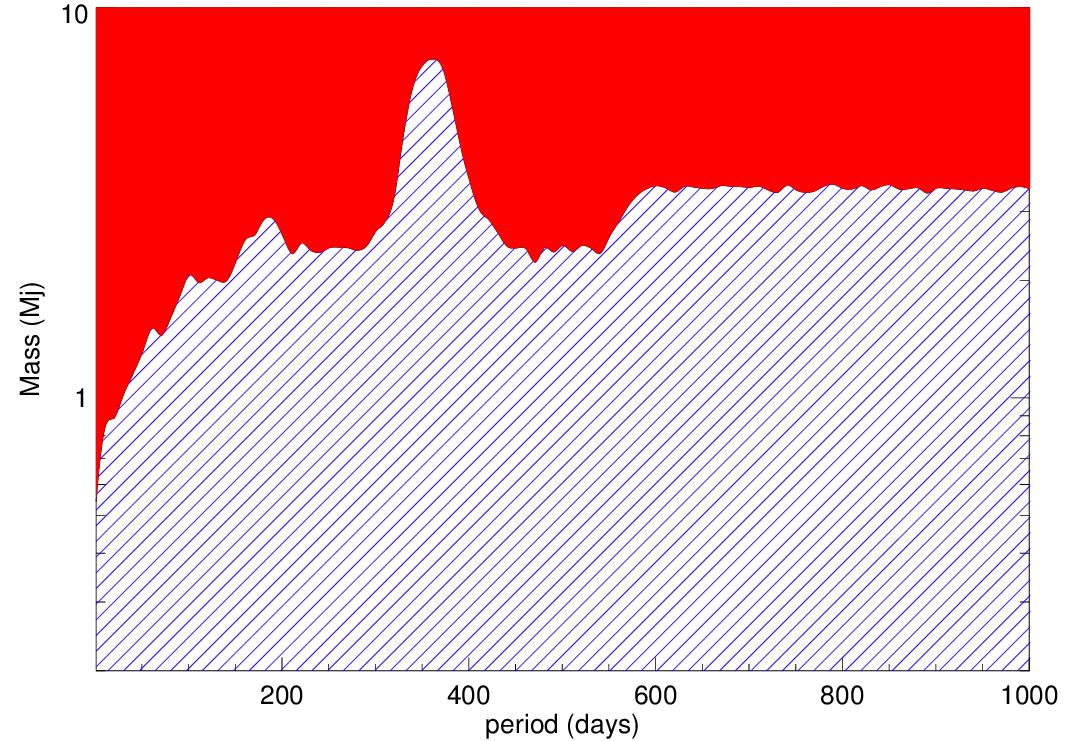}
 \caption{Planet detection regions for a star of the sample
 derived using the RV rms analysis. Red area: the rms of the simulated RV values
 is more than three times higher than the real measurement error, thus
 all potential planets can be detected. Blue ruled area: the rms of the simulated RV values
 is lower than three times the real measurement errors. Potential planets in this mass-period range
 cannot be detected with our observations.}
 \label{Examdeteffic}
 \end{figure}

 In order to take into account the contribution of stars with
 poor sampling and few observations, 
 we derived the number of detectable planets using the measurements rms.
 We considered the same grid of period and log-spaced mass cells as described in Sect.~\ref{sec:Methodology}.
 In each cell of the mass-period grid, we evaluated the average rms 
 of the synthetic radial velocities for the simulated planets ($RMS_{sim}$).
 We set our detection threshold by identifying
 for each star the mass-period cells with synthetic average
 RV variation more than three times higher than the estimated
 errors ($RMS_{sim} > 3\sigma_{corr}$) of our real measurements.
 Inflated measurements errors were used in the analysis.\\
 All potential planets with $RMS_{sim}$ smaller than 3$\sigma_{corr}$ 
 were instead considered undetectable.\\
 Finally, we folded the results of all the stars by 
 obtaining the mass-period cell-matrix $M$.
 The value for each $M_{ij}$ cell represents
 the total number of poorly sampled
 stars for which planets in that cell would be detected.\\
 As an example, the detection regions for a star of the sample
 derived using the RV rms analysis is shown in Fig.~\ref{Examdeteffic}.
 In the red area the rms of the simulated RV values ($RMS_{sim}$)
 is more than three times higher than the real measurements error ($\sigma_{corr}$):
 all hypothetical planets in this mass-period range would be detected.
 In the blue ruled area the rms of the simulated RV values ($RMS_{sim}$)
 is lower than $3\sigma_{corr}$:
 potential planets in this mass-period range cannot be detected with our observations.

 \subsubsection*{ \textit{Planetary occurrence}}

 We computed the planet occurrence rate $\gamma(P,M)$, 
 that is, the fraction of stars of our survey that are orbited by giant planets 
 in the selected period-mass ranges, using the following formalism.
 Considering the 24 stars analyzed with the Lomb-Scargle method,
 we derived the completeness, expressed as detectable planetary number, $R_{\star,k}$, for each $k$ star.
 First we weighted the calculated detection efficiency with the mass-period distribution of planets derived 
 from the relation of \citet{Cumming2008}: $df \propto M^{-0.31\pm0.2} P^{0.26\pm0.1} dlogM dlogP$. 
 Therefore we defined
 \begin{equation}
 R_{\star,k}=c \sum_{ij}^{Nbins} E_{ij}\cdot D_{ij}
,\end{equation}
\
\par\noindent
 where the sum is evaluated over the entire mass-period bins.
 $E_{ij}$ corresponds to the detection efficiency for each period-mass cell,
 $D_{ij}$ represents the mass-period distribution of planets within each bin and is calculated by
\begin{equation}
 D_{ij}=\int_{bin} M^{-0.31\pm0.2} P^{0.26\pm0.1} dlogM dlogP
.\end{equation}
\
\par\noindent
The normalization constant $c$ is defined conveniently inside our mass-period domain by the relation
\begin{equation}
 1=c\int_{1.0}^{1000}\int_{0.2}^{10} M^{-0.31\pm0.2} P^{0.26\pm0.1} dlogM dlogP
.\end{equation}
\
\par\noindent

Finally, we summed the values obtained from the analysis of all the MS and giant stars (G) as
 \begin{equation}
 R_{MStot}=\sum_{MS stars} R_{\star k},~~~~R_{Gtot}=\sum_{G stars} R_{\star k}.\end{equation}
\
\par\noindent
$R_{MStot}$ and $R_{Gtot}$ can be interpreted as the number of planets that would be detected
if every star in the sample were hosting a planet.
We imposed the result to be proportional to the number of real detected planets in the seven-year-long RV survey (three MS and two giant stars):
  \begin{equation}
 3=\gamma_{MS} R_{MStot},~~~~ 2=\gamma_{G} R_{Gtot}.
\end{equation}
\
\par\noindent
 Therefore, the proportional constants $\gamma_{MS}$ and $\gamma_{G}$ represent an estimate
 of the planet occurrence rate of our survey for the MS and G stars respectively.
 In the analysis, turn-off stars with 12.5$\leq M_{V}\leq$13.5 and 0.540$\leq(B-V)\leq$0.640 were considered 
 part of the MS sample, and the others were considere part of the G sample.
 
 Subsequently, we decided to include a correction for the total detection efficiency
 to take
 the contribution of the stars with few observations and poor sampling into account in computing the planet occurrence
rate.\\ 
 Two cases were evaluated.\\ 
 In the first, the planet occurrence among the MS stars (\textbf{N$_{MS}=14$}) 
 and the evolved stars (N$_{G}=10$) was rescaled
 with the ratio of N$_{MS}$ or N$_{G}$ 
 over the total number of stars considered in the analysis: 14
of 56 for the MS and 10 of 20 for the giant stars. 
 This corresponds to the more conservative approach, where all the stars with 
 few observations do not have a companion.\\
 In the second case, we instead
decided to rescale the detection efficiency for each cell $E_{ij}$ as
 \begin{equation}
 E_{ij}^{'}=\left(\frac{M_{ij}+N_{stars}}{N_{stars}}\right)\cdot E_{ij}
 ,\end{equation}
 \
 \par\noindent
 where $M_{ij}$ is the number of stars with potentially detectable planets for each mass-period cell 
 calculated using the method based on the RV value rms. In this case, only stars 
 with few observations and poor sampling were considered.
 $N_{stars}$ is the number of the stars with potentially detectable planets
 obtained with the periodogram-based analysis.

 \subsection{Discussion}

 When we consider only the 14 MS and the 10 evolved stars described in the previous section, 
 the planet occurrence in our survey 
 results in the values of $\gamma_{MS}\sim$58.3$^{+56.4}_{-31.1}$\% for the MS and $\gamma_{G}\sim$43.8$^{+57.2}_{-28.5}$\% for giants.
 The errors on the number of found planets are calculated following the prescription of \citet{Gehrels1986}.
 Combining the values for MS and giants gives a total occurrence rate of $\gamma_{tot}\sim$51.5$^{+35.0}_{-22.7}$\%.
 These 24 objects represent the fraction of the stars with a
detection efficiency high enough for the number of observations 
 to observe giant planets in the selected range of mass and period (see Figs.~\ref{PlEffMS} and~\ref{PlEffGTO}).\\
 Applying then the correction for the detectability
 based on those stars that have fewer observations,
 we obtained the following values for the two considered approaches. 
 In the more conservative case, without further detections,
 the planet occurrence becomes  
 $\sim$14.6$^{+14.1}_{-8.0}$\% for the MS stars, $\sim$21.9$^{+28.6}_{-14.2}$\% for the giant stars,
 and $\sim$16.3$^{+11.0}_{-7.0}$\% for the whole sample.\\  
 In the more general case,
 the procedure previously described instead gives a giant planet occurrence of 
 $\sim$15.8$^{+15.3}_{-8.4}$\% for the MS stars and $\sim$23.0$^{+29.9}_{-15.0}$\% for the evolved stars.
 When referring to the whole sample, the occurrence is $\sim$18.0$^{+12.0}_{-8.0}$\%.\\
 The values from the latter more general analysis are finally considered 
 as the final results of our study.
 Interestingly, they are 
 similar to the results of
 RV surveys around FGK field stars 
 that show exoplanet rates of $\sim$13\%
 for Jupiter-mass stars in approximately the same range of periods
 \citep{Mayor2011,Cumming2008}.\\
 Moreover, when we investigated only the 
 frequency of hot Jupiters around MS stars of our survey
and exactly repeated the
 procedure described in the previous sections, but focusing on objects 
 with periods shorter than 10 days,
 we found a hot Jupiter occurrence rate of $\sim$5.1$^{+4.9}_{-3.0}$\% for the conservative approach and
 of $\sim$5.7$^{+5.5}_{-3.0}$\% for the more general case. These values appear to be higher than 
 what is observed in field stars \citep{Wright2012,Howard11}.
 About this argument in particular, we refer to the discussion presented in our parallel work
 \citep{Brucalassi2016}. We pointed out in that paper that
 the high rate of hot Jupiters could be favored by a formation scenario dominated by strong encounters with other 
 stars or binary companions and subsequent planet-planet scattering, as predicted by N-body simulations.
 In this context, we should consider that rescaling the actual M67 detectability with the mass-period distribution
 derived by \citet{Cumming2008} for the field stars as done in the previous section could produce a biased result.
 However, future investigations and a larger number of planet detections 
 in stars cluster are required to assess this hypothesis.
 
 Finally, it is worth to point out that we did not find
 any clear detection of long-period giant planets around MS stars,
 possibly because the sensitivity of our survey decreases heavily for long-period planets.
 However, the recent discovery in the Praesepe cluster of a second massive outer planet around Pr0211 
 that hosts a close hot Jupiter on
a slightly eccentric orbit \citep{Malavolta2016} suggests
 that further long-term monitoring could yield indications of interesting discoveries in OCs.

\section{Summary and conclusion}   
\label{sec:Conclusion}

 We have presented the results of a long-term search program for giant planets in the solar-age, 
 solar-metallicity open cluster M67.

 Five different instruments were used, and after finding proper zero-point corrections to HARPS, 
 1145 observations for 88 stars were analyzed.\\
 The problem of combining long-term precision RV data from different instruments complicates the analysis.
 Zero-points offsets were derived with a limited precision, which leads to a loss of sensitivity for trends and long periods.
 However, this problem quite frequently occurs in long-term surveys
because spectrographs receive upgrades or survey projects are transferred
 to new instruments. 
 Long-term access to the same telescope or instrument configuration is therefore quite important
 for this type of studies.
 
 Five stars of our sample are found to host planets.
 One of them (YBP401) has recently been presented in \citet{Brucalassi2016}, and in that work
 we have refined the parameters, based on new measurements, 
 of other two planets announced previously around YBP1194 and YBP1514 \citep{Brucalassi2014}.
 The other planet-host candidate S978 was discussed in this paper.
 We have no clear additional planet detection in our sample, but some promising or controversial cases such as  
 YBP778 and YBP2018 call for follow-up observations.
 Fourteen new binaries were identified and added to the catalog of known M67 binaries 
 that has been created in Paper I \citep{Pasquini2012}. 
 Moreover, we see trends in S1062 and S1137, and large RV scattering is exhibited by S815 and S488, 
 which might be explained by stellar companions or substellar objects.
 However, stellar chromospherical activity and magnetic cycles may cause RV variations that might be mistaken as companions.
 Activity indicators such as the Ca II H$\&$K lines, H$\alpha$, and the bisector span are fundamental to verify 
 claimed planets from RV variability.
 
 We provided an estimate of the planet occurrence rate $\gamma(P,M)$, 
 that is, the fraction of stars of our survey that is orbited by Jupiter planets 
 in the ranges of period between 1.0 day and 1000 days and has
a planet mass between 0.2\JM and 10.0\JM.
 Although one of the main problems of this survey was the poor 
 and sparse sampling and the small number of observations per star,
 which complicates the statistical analysis and increases the uncertainty 
 on the frequency of Jupiters in our survey,  
 we find a total occurrence rate of Jupiter-mass planets in the M67 sample of 
 $\sim$18.0$^{+12.0}_{-8.0}$\% (precisely $\sim$15.8$^{+15.3}_{-8.4}$\% for the MS stars 
 and $\sim$23.0$^{+29.9}_{-15.0}$\% for the evolved stars).
 It is worth noting that on average we need a minimum number of 20 observations per star 
 to exclude Jupiter-mass planets at high confidence when a periodogram analysis is applied.\\
 The Jupiter-mass planets in our sample imply that our results are consistent with the planet frequency 
 of much larger surveys carried out for field star samples. For example, 
 from the ELODIE survey, \citet{Naef2005}
 estimated that a fraction of 7.5$\pm$1.5\% stars host giant planets with periods shorter than 10 yr, while 
 \citet{Cumming2008}
 derived a frequency of 12$\pm$1.6\% from the Keck survey and \citet{Mayor2011} 13.9$\pm$1.7\% from the HARPS/CORALIE survey.\\
 As a general conclusion, our simulation study seems to confirm the recent finding \citep{Meibom2013}
 that the frequency of massive planets around stars of open clusters agrees 
 with the frequency of these planets around field stars.\\
 However, we note that when we only investigate the 
 frequency of hot Jupiters around MS stars of our survey, we find a hot Jupiter occurrence rate ($\sim$5.7$^{+5.5}_{-3.0}$\%) 
 that is substantially higher than what is observed in field stars \citep[see also][]{Brucalassi2016}. 
 
 We have shown that the search for planets in open clusters is a powerful test benchmark 
 for the theory of planet formation and stellar evolution. 
 The new generation of spectrographs
 such as ESPRESSO \citep{Pasquini2009} will extend this search more effectively.
 It is clear, however, that to efficiently perform a planet
 search survey in star clusters, a multi-object capability,
 even with relatively low multiplex (10-50), is desirable.
 Unfortunately, no such facility is currently planned.

 \begin{acknowledgements}
 LP acknowledges the Visiting Researcher program of the CNPq Brazilian Agency, at the
 Fed. Univ. of Rio Grande do Norte, Brazil.
 RPS thanks ESO DGDF, the whole GAPS-TNG community, the Hobby Eberly Telescope (HET) project, the PNPS and PNP of INSU - CNRS 
 for allocating the observations and the technical support.
 MTR received support from PFB06 CATA (CONICYT).
 We are grateful to Gaspare Lo Curto and Michele Cappetta for the support in data reduction analysis 
 and helpful discussions.
% Part of this work was supported by the German
 %      \emph{Deut\-sche For\-schungs\-ge\-mein\-schaft, DFG\/} project
  %    number Ts~17/2--1.
\end{acknowledgements}

%_____________________________________________________________
%                              Table longer than a single page  
%-------------------------------------------------------------
% All long tables will be placed automatically at the end, after 
%                                        \end{thebibliography}
%
\begin{longtab}
\begin{longtable}{llllrrrrrrrrr}
\caption{\label{table:observations}Observed targets in M67. Object name, basic stellar parameters, number of observations
for each instrument (H: HARPS, S: SOPHIE, C: CORALE, HET: HRS at Het, H-N: HARPS-N), total number of observations,
mean stellar RV, average measurements error, average corrected measurements error, observed RV dispersion. Binary candidates are indicated in bold face.}\\
\hline\hline
%\footnotesize
Object& $B-V$ & $M_{V}$ & H & S & C & HET & H-N & TOT & RV(\kms) & $\sigma_{obs}$\small{(\kms)}& $\sigma_{cor}$\small{(\kms)}& $\sigma_{RV}$\small{(\kms)} \\
\hline
\endfirsthead
\caption{continued.}\\
\hline\hline
Object& $B-V$ & $M_{V}$ & H & S & C & HET & H-N & TOT & RV(\kms) & $\sigma_{obs}$\small{(\kms)}&$\sigma_{cor}$\small{(\kms)}& $\sigma_{RV}$\small{(\kms)} \\
\hline
\endhead
\hline
\endfoot
\textbf{YBP219}  & \textbf{0.570} & \textbf{13.6} &\textbf{6}&  \textbf{5} & \textbf{0}&  \textbf{1} & \textbf{0} & \textbf{11} & \textbf{33.714} & \textbf{0.016} &\textbf{0.022} &\textbf{0.323} \\
YBP266  & 0.570 & 13.6 & 11 &  3 &  0 &  0 &  1 & 15 & 33.774 & 0.017&0.022&0.031 \\
YBP285  & 0.663 & 14.5 &  7 &  2 &  0 &  0 &  0 &  9 & 34.397 & 0.016&0.022&0.019 \\
\textbf{YBP288}  & \textbf{0.637} & \textbf{13.9} & \textbf{1}&  \textbf{2} & \textbf{0}&  \textbf{0} & \textbf{0} & \textbf{3} & \textbf{37.691} & \textbf{0.010}&\textbf{0.018}&\textbf{1.299} \\
YBP291  & 0.570 & 13.5 & 23 &  3 &  0 &  0 &  0 & 26 & 32.521 & 0.017&0.022&0.026 \\     
YBP349  & 0.636 & 14.3 & 11 &  0 &  0 &  0 &  0 & 11 & 35.048 & 0.014&0.020&0.023 \\     
YBP350  & 0.561 & 13.6 & 11 &  2 &  0 &  0 &  0 & 13 & 33.227 & 0.015&0.021&0.020 \\     
YBP401  & 0.566 & 13.7 &  4 &  4 &  4 &  0 &  2 & 31 & 33.178 & 0.015&0.021&0.035 \\     
YBP473  & 0.658 & 14.4 &  7 &  0 &  0 &  0 &  0 &  7 & 33.266 & 0.016&0.022&0.023 \\     
YBP587  & 0.605 & 14.1 &  6 &  2 &  0 &  0 &  0 &  8 & 33.188 & 0.013&0.020&0.028 \\     
YBP613  & 0.612 & 13.3 & 10 &  4 &  0 &  0 &  0 & 14 & 33.565 & 0.015&0.021&0.020 \\     
YBP637  & 0.661 & 14.5 &  5 &  2 &  0 &  0 &  0 &  7 & 34.801 & 0.014&0.020&0.019 \\     
\textbf{YBP673}  & \textbf{0.665} & \textbf{14.4} & \textbf{14} &  \textbf{1} &  \textbf{0} & \textbf{14} & \textbf{2} & \textbf{31} & \textbf{33.496} & \textbf{0.026}&\textbf{0.030}&\textbf{0.249} \\
YBP689  & 0.622 & 13.1 & 10 &  2 &  0 &  3 &  0 & 15 & 33.650 & 0.017&0.022&0.044 \\   
YBP750  & 0.598 & 13.6 &  4 &  1 &  0 &  0 &  0 &  5 & 34.251 & 0.014&0.020&0.011 \\   
\textbf{YBP769}  & \textbf{0.665} & \textbf{14.4} & \textbf{2} &  \textbf{1} &  \textbf{0} & \textbf{0} & \textbf{0} & \textbf{3} & $CCFDouble peak$  &  & &\\
YBP778  & 0.582 & 13.1 & 15 &  4 &  0 &  0 &  2 & 21 & 34.288 &0.016&0.022&0.090 \\
YBP809  & 0.696 & 15.0 &  5 &  0 &  0 &  0 &  0 &  5 & 32.864 &0.011 &0.019&0.011 \\
\textbf{YBP851}  & \textbf{0.617} & \textbf{14.4} & \textbf{3} &  \textbf{1} &  \textbf{0} & \textbf{0} & \textbf{0} & \textbf{4} & \textbf{33.759} &  \textbf{0.010}&\textbf{0.018}&\textbf{1.417} \\
\textbf{YBP911}  & \textbf{0.673} & \textbf{14.6} & \textbf{2} &  \textbf{0} &  \textbf{0} & \textbf{0} & \textbf{0} & \textbf{2} & \textbf{33.738} &  \textbf{0.020}&\textbf{0.025}&\textbf{0.703} \\
YBP988  & 0.598 & 14.2 &  5 &  1 &  0 &  0 &  0 &  6 & 32.862 & 0.013&0.020&0.016 \\   
YBP1032 & 0.598 & 14.4 &  6 &  0 &  0 &  0 &  0 &  6 & 34.913 & 0.013&0.020&0.013 \\   
YBP1036 & 0.690 & 15.0 & 15 &  0 &  0 &  0 &  0 & 15 & 34.025 & 0.015&0.021&0.025 \\   
\textbf{YBP1051} & \textbf{0.595} &  \textbf{14.1} & \textbf{15} & \textbf{3} & \textbf{0} & \textbf{5} & \textbf{0} &  \textbf{23} &  \textbf{33.364} & \textbf{0.016} &\textbf{0.022} &\textbf{0.156} \\
YBP1062 & 0.626 & 14.5 & 19 &  3 &  0 &  0 &  0 & 22 & 33.462 & 0.017&0.022&0.067 \\  
\textbf{YBP1067}  & \textbf{0.642} & \textbf{14.6} & \textbf{7} &  \textbf{0} &  \textbf{0} & \textbf{2} & \textbf{0} & \textbf{9} & \textbf{33.667} &\textbf{0.019} &\textbf{0.024} &\textbf{1.030} \\
YBP1075 & 0.633 & 13.7 & 11 &  1 &  0 &  0 &  1 & 13 & 33.858 & 0.014&0.020&0.034 \\   
YBP1088 & 0.618 & 14.5 &  8 &  0 &  0 &  0 &  0 &  8 & 33.434 & 0.014&0.020&0.022 \\   
\textbf{YBP1090}  & \textbf{0.650} & \textbf{13.8} & \textbf{2} &  \textbf{1} &  \textbf{0} & \textbf{0} & \textbf{0} & \textbf{3} & \textbf{35.186} & \textbf{0.021}&\textbf{0.025}&\textbf{1.265} \\
YBP1101 & 0.661 & 14.7 &  4 &  1 &  0 &  0 &  0 &  5 & 33.484 & 0.015&0.021&0.029 \\
YBP1129 & 0.583 & 14.2 &  4 &  1 &  0 &  0 &  0 &  5 & 34.479 & 0.009&0.018&0.009 \\
YBP1137 & 0.657 & 14.9 & 13 &  0 &  0 &  0 &  0 & 13 & 34.227 & 0.018&0.023&0.031 \\
YBP1194 & 0.626 & 14.6 & 14 &  3 &  0 &  5 &  2 & 26 & 34.189 & 0.017&0.022&0.027 \\
YBP1197 & 0.565 & 13.3 & 10 &  0 &  0 &  0 &  0 & 10 & 34.591 & 0.017&0.022&0.015 \\
YBP1247 & 0.568 & 14.1 &  5 &  1 &  0 &  0 &  0 &  6 & 32.966 & 0.013&0.020&0.014 \\
YBP1303 & 0.636 & 14.6 &  4 &  1 &  0 &  0 &  0 &  5 & 33.395 & 0.017&0.022&0.019 \\
\textbf{YBP1304} & \textbf{0.723} & \textbf{14.7} & \textbf{2} &  \textbf{1} &  \textbf{0} & \textbf{0} & \textbf{0} & \textbf{3} & \textbf{32.512} &\textbf{0.019} &\textbf{0.024} &\textbf{2.670} \\
\textbf{YBP1315} & \textbf{0.693} & \textbf{14.3} & \textbf{5} &  \textbf{2} &  \textbf{0} & \textbf{6} & \textbf{0} & \textbf{13} & \textbf{34.885} &\textbf{0.022} &\textbf{0.026}&\textbf{0.801} \\
YBP1334 & 0.639 & 14.4 &  6 &  1 &  0 &  0 &  0 &  7 & 33.066 & 0.014&0.020&0.021 \\
YBP1387 & 0.585 & 14.1 &  5 &  1 &  0 &  0 &  0 &  6 & 34.059 & 0.012&0.019&0.017 \\
YBP1392 & 0.675 & 14.8 &  8 &  1 &  0 &  0 &  0 &  6 & 34.556 & 0.016&0.022&0.023 \\
YBP1458 & 0.698 & 15.0 &  6 &  0 &  0 &  0 &  0 &  6 & 33.417 & 0.013&0.020&0.015 \\
YBP1496 & 0.556 & 13.9 &  8 &  1 &  0 &  0 &  0 &  9 & 34.786 & 0.015&0.021&0.017 \\
YBP1504 & 0.584 & 14.2 &  4 &  1 &  0 &  0 &  0 &  5 & 33.761 & 0.013&0.020&0.018 \\
YBP1514 & 0.680 & 14.8 &  8 &  8 &  8 &  8 &  8 & 25 & 34.048 & 0.018&0.023&0.036 \\
YBP1587 & 0.600 & 14.2 & 11 &  1 &  0 &  0 &  1 & 13 & 33.446 & 0.011&0.019&0.035 \\
YBP1622 & 0.591 & 14.2 &  6 &  0 &  0 &  0 &  0 &  6 & 33.951 & 0.012&0.019&0.015 \\
\textbf{YBP1716} & \textbf{0.619} & \textbf{13.3} & \textbf{5} &  \textbf{1} &  \textbf{0} & \textbf{7} & \textbf{0} & \textbf{13} & \textbf{36.205} &\textbf{0.010} &\textbf{0.018} &\textbf{0.651} \\
YBP1722 & 0.560 & 14.1 & 10 &  1 &  0 &  5 &  1 & 17 & 34.483 & 0.018&0.023&0.025 \\  
YBP1735 & 0.620 & 14.3 &  5 &  0 &  0 &  0 &  0 &  5 & 33.976 &0.013 &0.020&0.014 \\  
\textbf{YBP1758}  & \textbf{0.653} & \textbf{13.2} & \textbf{2} &  \textbf{1} &  \textbf{0} & \textbf{0} & \textbf{0} & \textbf{3} & \textbf{29.653} &\textbf{0.011} &\textbf{0.019} &\textbf{1.521} \\
YBP1768 & 0.615 & 14.4 &  3 &  0 &  0 &  0 &  0 &  3 & 34.497 & 0.012&0.019&0.003 \\
YBP1787 & 0.626 & 14.5 &  8 &  1 &  0 &  0 &  0 &  9 & 34.065 & 0.014&0.020&0.016 \\
YBP1788 & 0.622 & 14.4 &  8 &  0 &  0 &  0 &  0 &  8 & 34.162 & 0.014&0.020&0.042 \\
YBP1852 & 0.572 & 14.0 &  7 &  1 &  0 &  0 &  0 &  8 & 32.914 & 0.013&0.020&0.018 \\
YBP1903 & 0.648 & 14.7 &  5 &  0 &  0 &  0 &  0 &  5 & 33.390 & 0.017&0.022&0.023 \\
YBP1948 & 0.571 & 14.0 &  4 &  1 &  0 &  0 &  0 &  5 & 33.333 & 0.012&0.019&0.011 \\
YBP1955 & 0.589 & 14.2 &  6 &  1 &  0 &  0 &  0 &  7 & 33.212 & 0.013&0.020&0.021 \\
YBP2018 & 0.631 & 14.6 & 22 &  2 &  0 &  7 &  0 & 31 & 31.991 & 0.020&0.025&0.089 \\
S364    & 1.360 &  9.8 & 14 &  6 &  7 &  5 &  0 & 32 & 33.198 & 0.010&0.018&0.042 \\
S488    & 1.550 &  8.9 & 15 & 11 &  0 &  8 &  3 & 37 & 32.861 & 0.012&0.019&0.156 \\
S602    & 0.512 & 12.9 &  6 &  4 &  0 &  7 &  0 & 17 & 33.894 & 0.018&0.023&0.047 \\       
S610    & 0.493 & 12.9 &  6 &  1 &  0 &  5 &  0 & 11 & 33.371 & 0.024&0.027&0.056 \\       
S657    & 0.559 & 12.3 &  8 &  0 &  0 &  0 &  0 &  8 & 33.234 & 0.008&0.017&0.015 \\         
S731    & 0.516 & 13.1 & 10 &  1 &  0 &  0 &  0 & 11 & 33.097 & 0.012&0.019&0.021 \\         
S815    & 0.497 & 12.9 & 14 &  4 &  0 &  7 &  3 & 28 & 34.139 & 0.019&0.024&0.207 \\         
S978    & 1.332 &  9.7 & 16 &  9 & 10 &  3 &  2 & 40 & 34.577 & 0.010&0.018&0.043 \\         
S989    & 1.048 & 11.4 &  8 &  1 &  8 &  3 &  0 & 20 & 34.795 & 0.019&0.024&0.024 \\         
S1001   & 0.759 & 12.4 &  3 &  1 &  0 &  0 &  0 &  4 & 33.409 & 0.007&0.017&0.024 \\         
S1010   & 1.069 & 10.5 &  7 &  3 &  6 &  0 &  0 & 16 & 33.746 & 0.009&0.018&0.025 \\ 
S1016   & 1.098 & 10.3 &  7 & 11 &  0 &  0 &  0 & 18 & 33.872 & 0.004&0.016&0.072 \\ 
S1054   & 0.859 & 11.2 &  8 &  1 &  9 &  0 &  0 & 18 & 33.500 & 0.015&0.021&0.025 \\ 
S1074   & 1.111 & 10.4 &  3 &  1 &  9 &  0 &  0 & 13 & 34.139 & 0.014&0.020&0.042 \\ 
S1084   & 1.086 & 10.5 & 15 &  1 &  6 &  5 &  0 & 27 & 33.900 & 0.007&0.017&0.021 \\ 
S1230   & 0.524 & 13.1 &  4 &  0 &  0 &  0 &  0 &  3 & 33.773 & 0.013&0.020&0.005 \\ 
S1254   & 0.999 & 11.5 & 10 &  3 &  5 &  0 &  0 & 18 & 32.867 & 0.013&0.020&0.058 \\ 
S1271   & 0.506 & 12.9 & 11 &  2 &  0 &  0 &  0 & 13 & 33.648 & 0.016&0.022&0.023 \\ 
S1279   & 1.081 & 10.6 &  7 &  2 &  7 &  0 &  0 & 16 & 33.384 & 0.010&0.018&0.021 \\ 
S1288   & 1.016 & 11.3 &  5 &  1 &  6 &  0 &  0 & 12 & 33.454 & 0.020&0.025&0.040 \\ 
S1293   & 0.565 & 12.1 & 13 & 10 &  0 &  7 &  0 & 30 & 34.094 & 0.010&0.018&0.042 \\  
S1305   & 0.945 & 12.2 & 14 &  2 &  0 &  6 &  0 & 22 & 33.964 & 0.011&0.019&0.020 \\  
S1316   & 1.077 & 10.6 &  9 &  2 &  6 &  4 &  0 & 21 & 32.860 & 0.012&0.019&0.028 \\  
S1402   & 1.109 & 10.9 &  4 &  1 &  0 &  0 &  0 &  5 & 33.781 & 0.007&0.017&0.008 \\  
S1479   & 0.682 & 10.5 &  6 &  1 &  6 &  0 &  0 & 13 & 34.319 & 0.011&0.019&0.013 \\  
S1557   & 1.249 & 10.1 & 13 & 13 &  6 &  4 &  1 & 37 & 33.797 & 0.010&0.018&0.138 \\  
\textbf{S1583}  & \textbf{0.665} & \textbf{14.4} & \textbf{2} &  \textbf{1} &  \textbf{0} & \textbf{0} & \textbf{0} & \textbf{2} & $CCFDouble peak$  &  & &\\
S1592   & 1.032 & 10.5 &  4 &  2 &  8 &  0 &  0 & 14 & 33.639 & 0.011&0.018&0.024 \\
S1607   & 0.548 & 12.7 & 12 &  6 &  0 &  2 &  2 & 22 & 33.391 & 0.013&0.019&0.062 \\

\end{longtable}
\end{longtab}

%-------------------------------------------------------------------

\bibliographystyle{aa} % style aa.bst
%\bibliography{ABRUCALASSI} % your references Yourfile.bib
\bibliography{master} % your references Yourfile.bib

%-------------------------------------------------------------
\newpage

\Online

\begin{appendix} %First online appendix
\section{Accompanying Plots}

\begin{figure}[!h]
%\centering
%\includegraphics[width=16.4cm,clip]{./Figures/PrmsContour1_S1607.jpg}
     \resizebox{1.90\hsize}{!}
            {
            \includegraphics{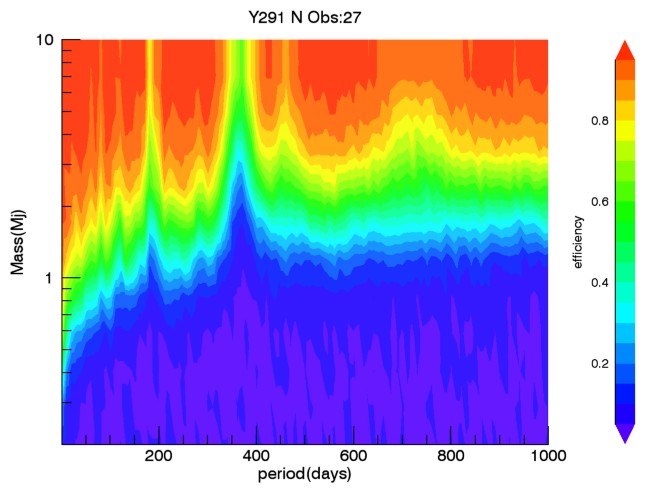}
            \includegraphics{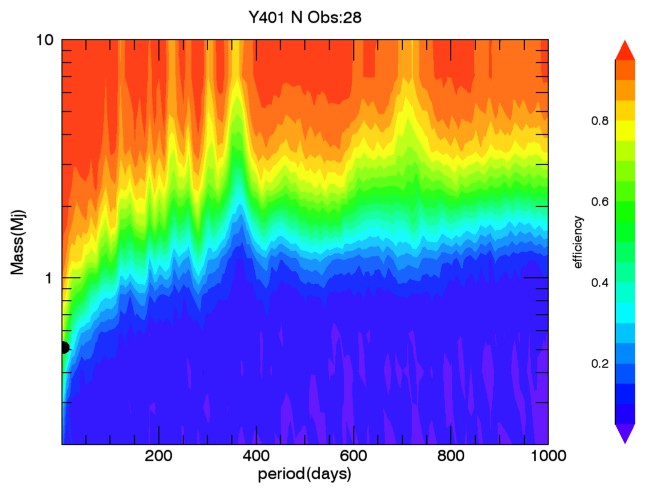}
            \includegraphics{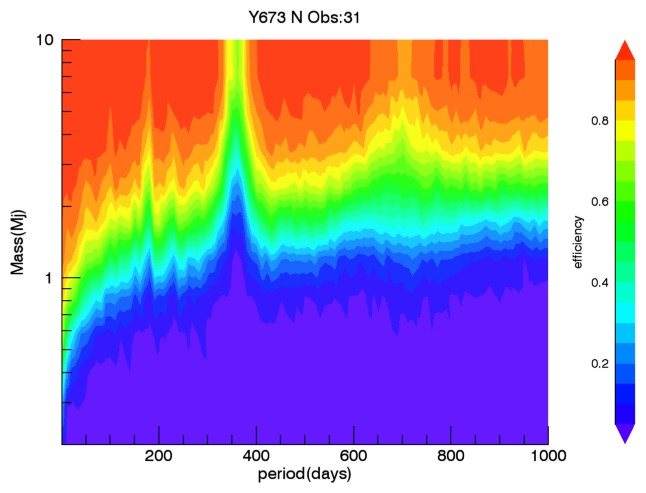}
            }
      \resizebox{1.90\hsize}{!}      
             {      
            \includegraphics{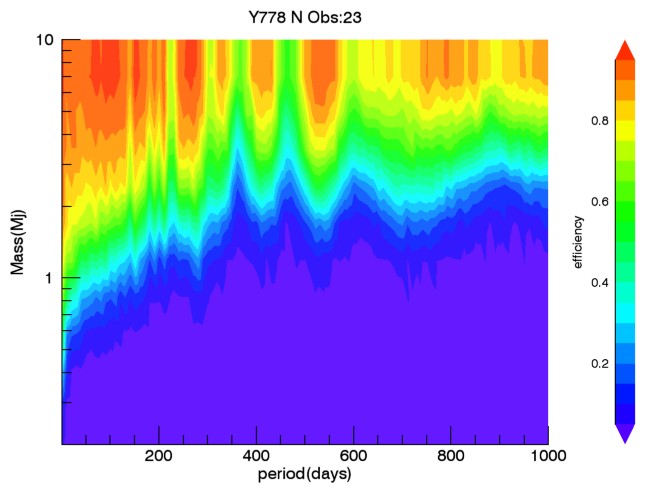}
            \includegraphics{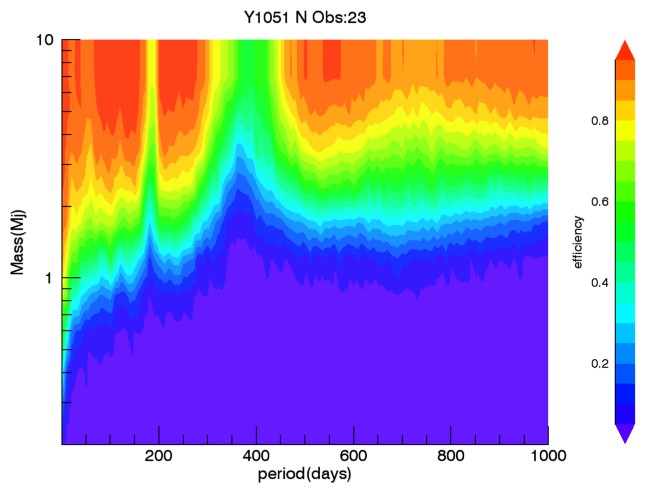}
            \includegraphics{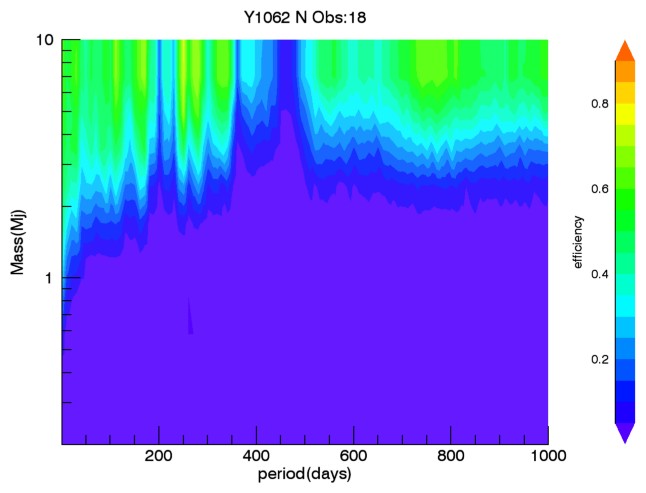}
          }
        \resizebox{1.90\hsize}{!}  
               {  
            \includegraphics{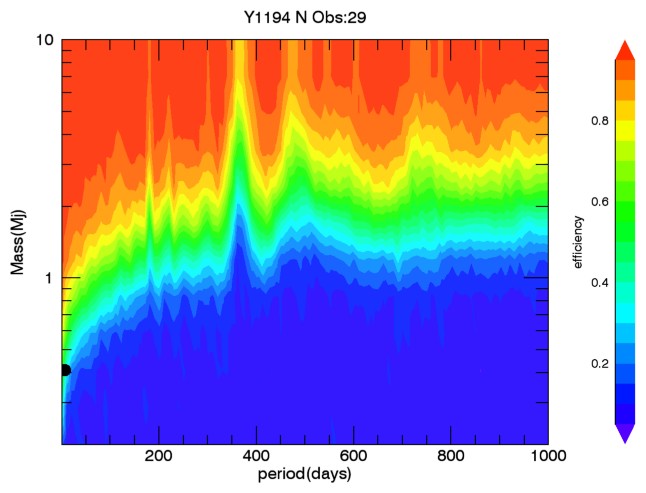}
            \includegraphics{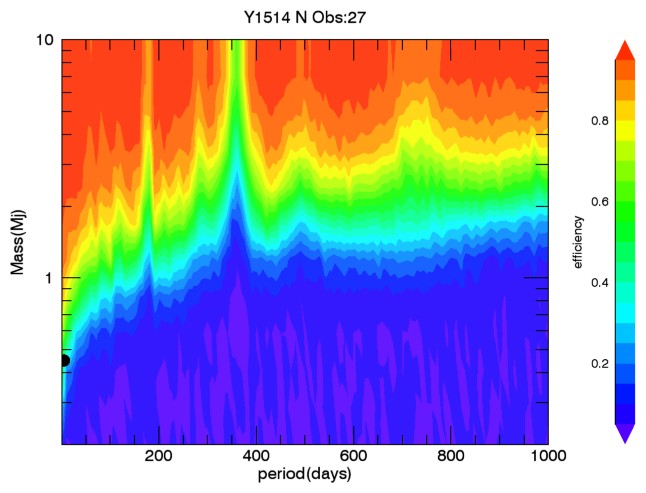}
            \includegraphics{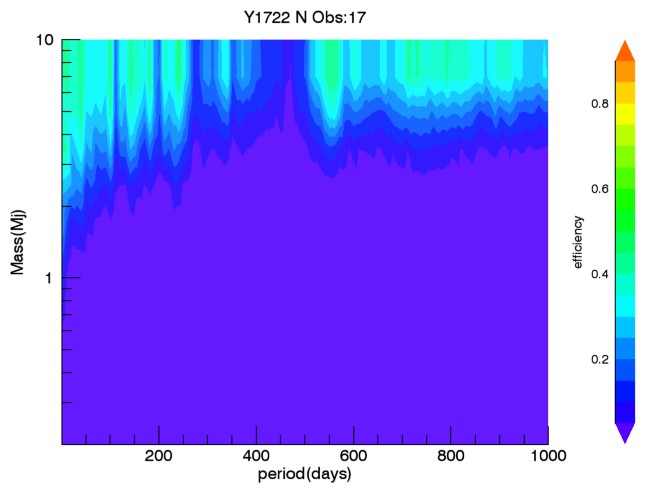}
          }
       \resizebox{1.90\hsize}{!}   
              {   
            \includegraphics{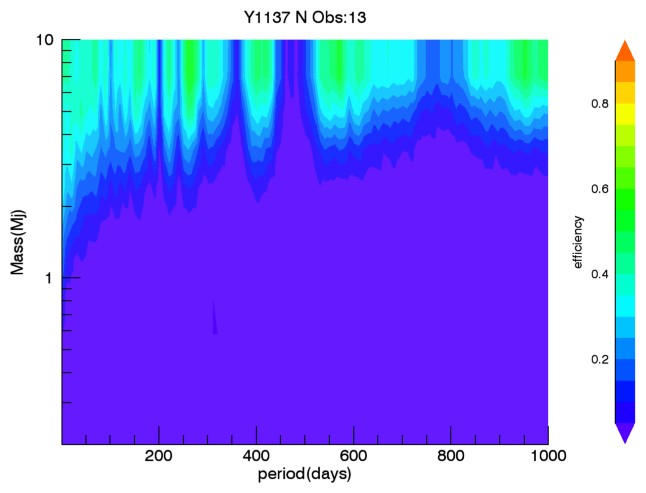} 
            \includegraphics{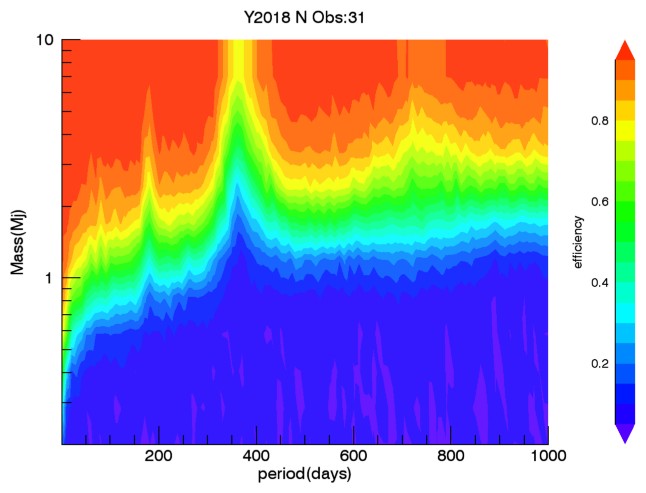}
            \includegraphics{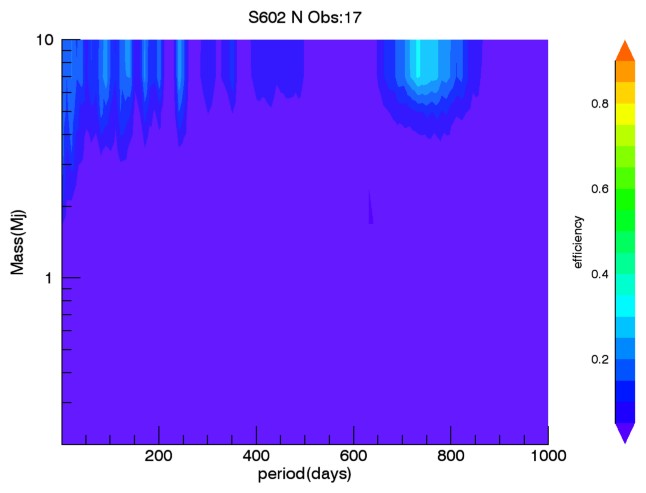} 
           }
         \resizebox{1.27\hsize}{!}  
                { 
            \includegraphics{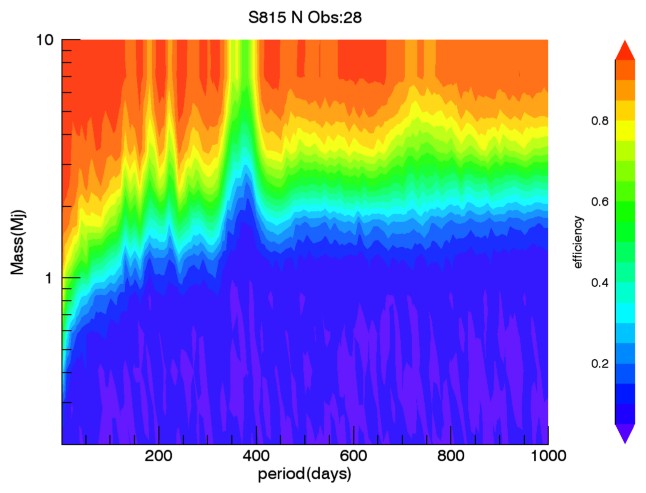}
            \includegraphics{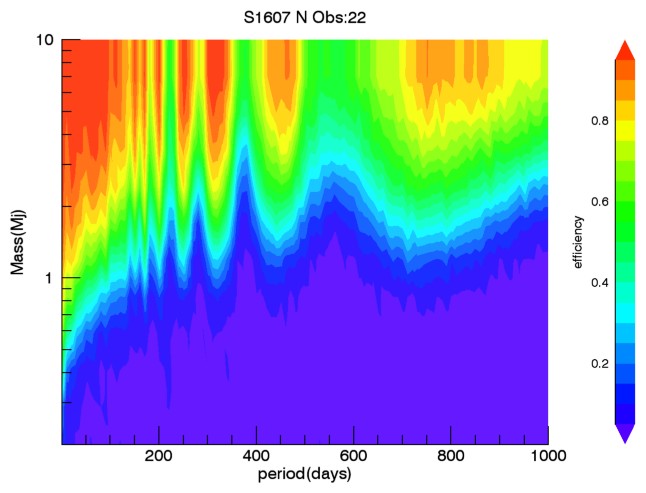}
           } 
      \begin{minipage}{\textwidth}    
      \caption{ \label{PlEffMS} Contours of the planet detection efficiency for the MS stars in the domains of 1-1000 days for planet period            
       and 0.2-10 \JM \ for planet mass. A 0.01 FAP level has been used as detection planet threshold in the periodogram analysis. 
       A filled circle indicates the planet position.} 
        \end{minipage}

\end{figure}

\newpage

\begin{figure*}
\begin{center}
        \includegraphics[scale=0.25]{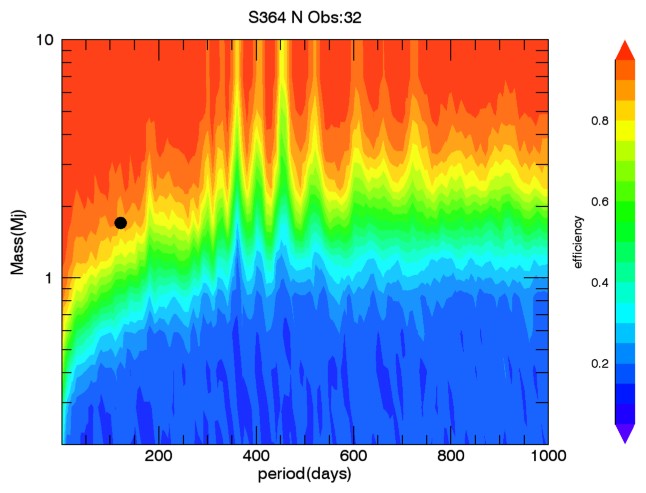} 
       \includegraphics[scale=0.25]{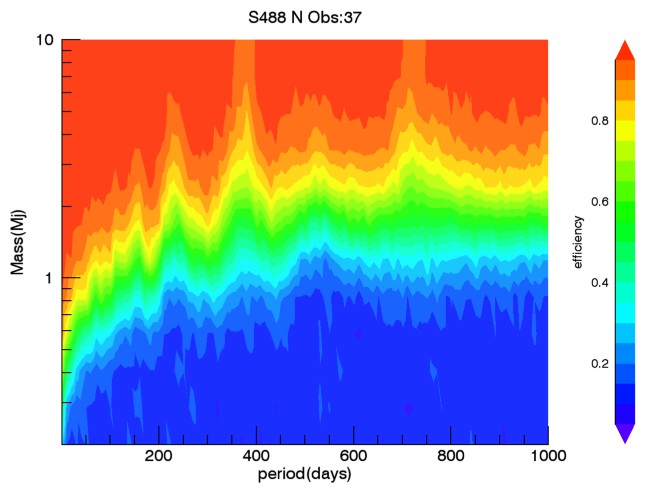}
       \includegraphics[scale=0.25]{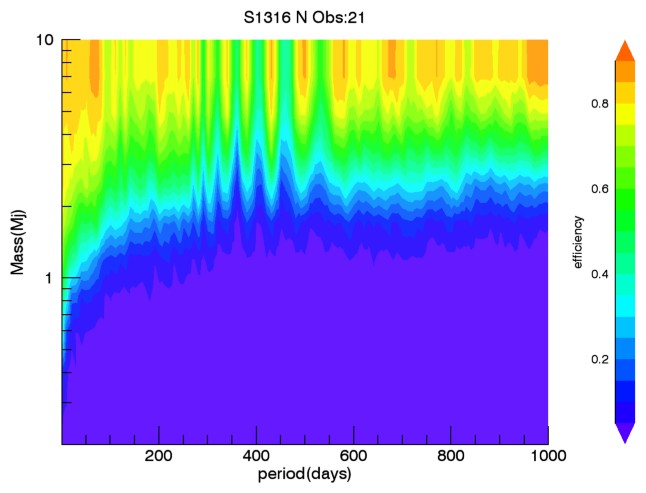}
       \includegraphics[scale=0.25]{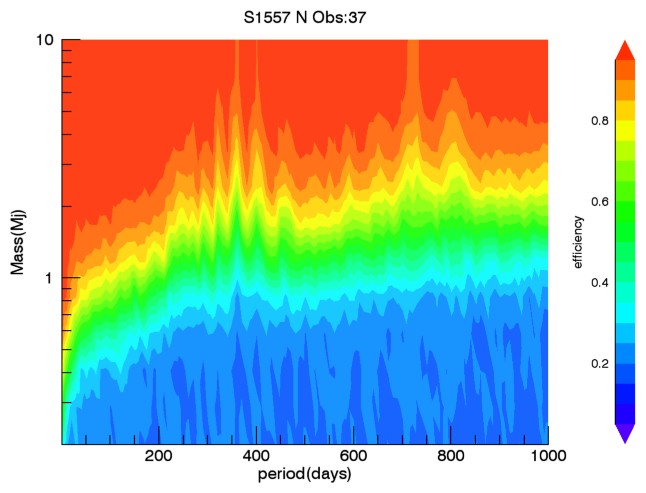}
      \includegraphics[scale=0.25]{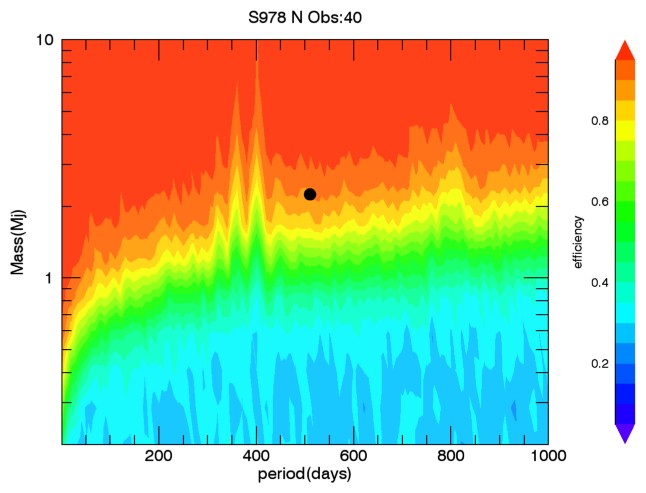}
      \includegraphics[scale=0.25]{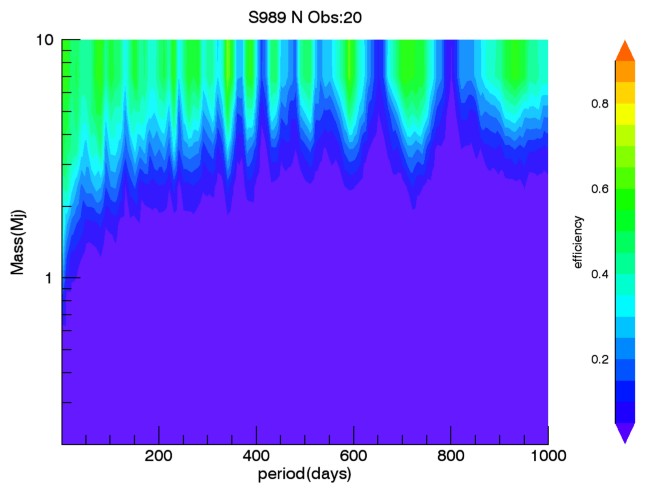}
      \includegraphics[scale=0.25]{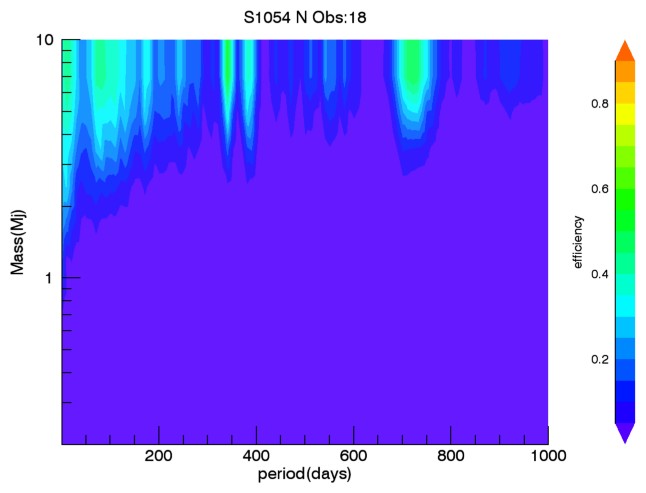}
      \includegraphics[scale=0.25]{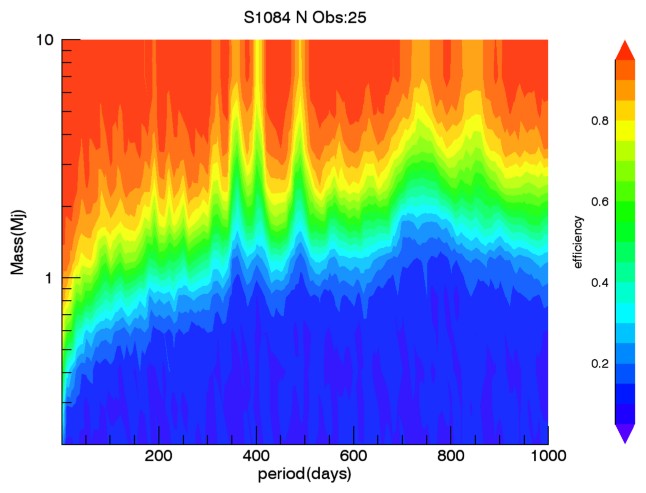}
      \includegraphics[scale=0.25]{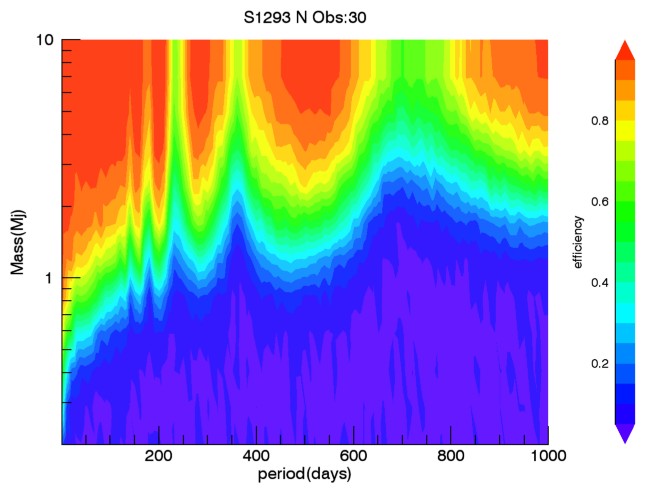}
      \includegraphics[scale=0.25]{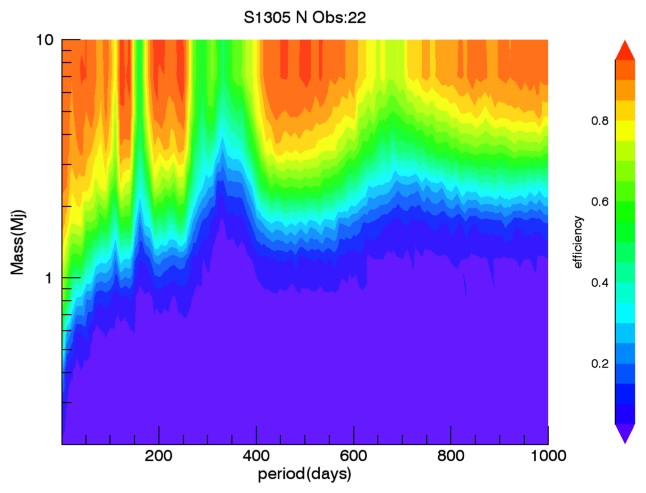}
 \end{center}  
 \caption{Contours of the planet detection efficiency for the evolved stars in the domains of 1-1000 days for planet period            
       and 0.2-10 \JM \ for planet mass. A 0.01 FAP level has been used as detection planet threshold in the periodogram analysis. 
       A filled circle indicates the planet position.}  
 
 \label{PlEffGTO}
\end{figure*}

\newpage

\begin{figure*}[ht]
                \begin{center}
                        \begin{tabular}{c}
                               \hspace{.4cm}
                                RV values \hspace{3.0cm} Residuals \hspace{3.0cm} BIS span \hspace{3.0cm} FWHM\\

  \resizebox{\hsize}{!}
       {
   \includegraphics{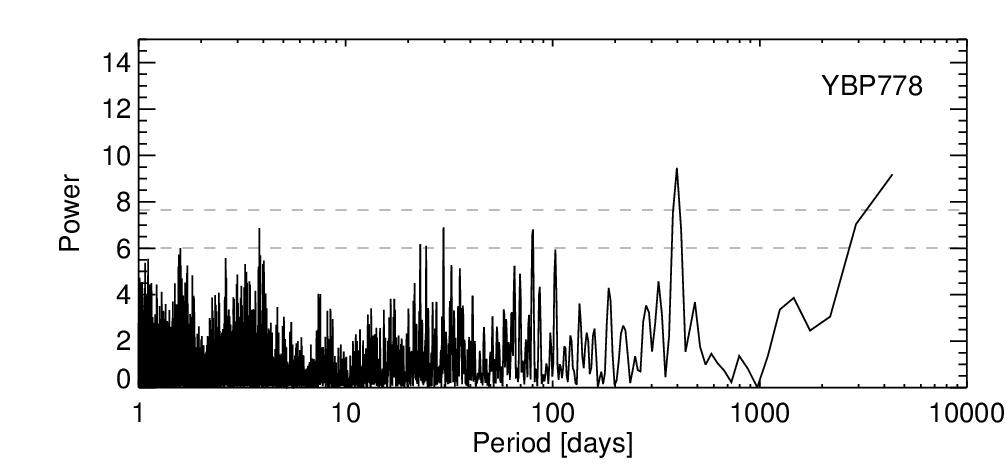}
    \includegraphics{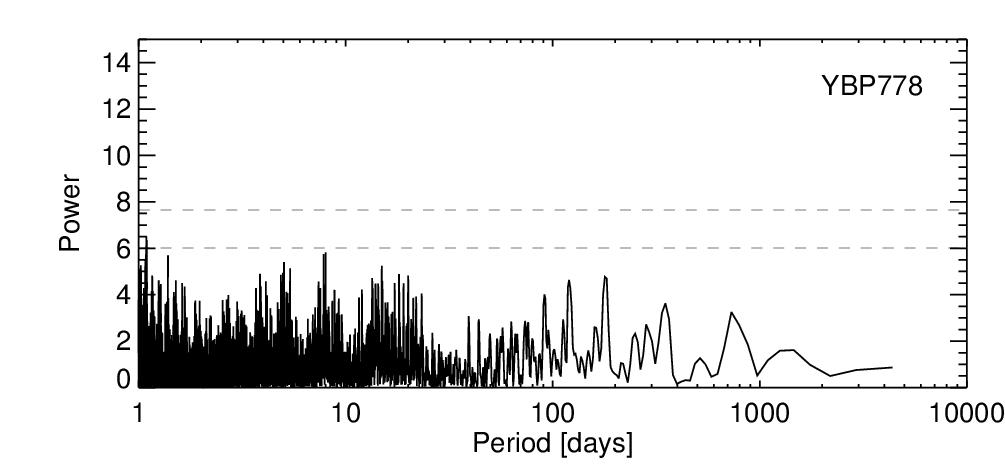}
    \includegraphics{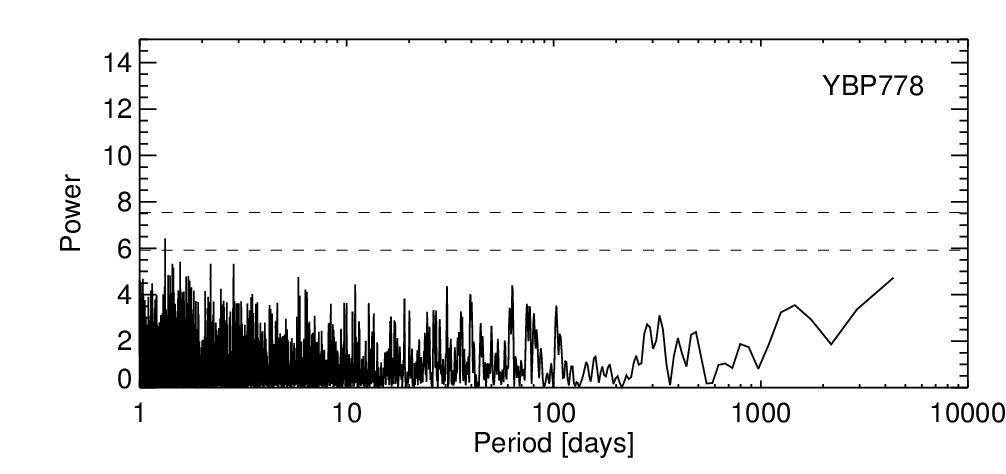} 
    \includegraphics{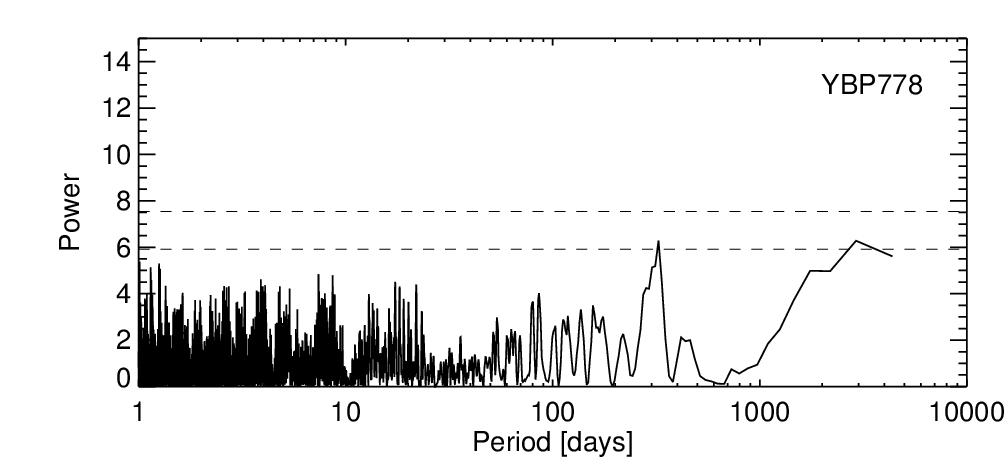}  
       }\\
    \resizebox{\hsize}{!}
       {
   \includegraphics{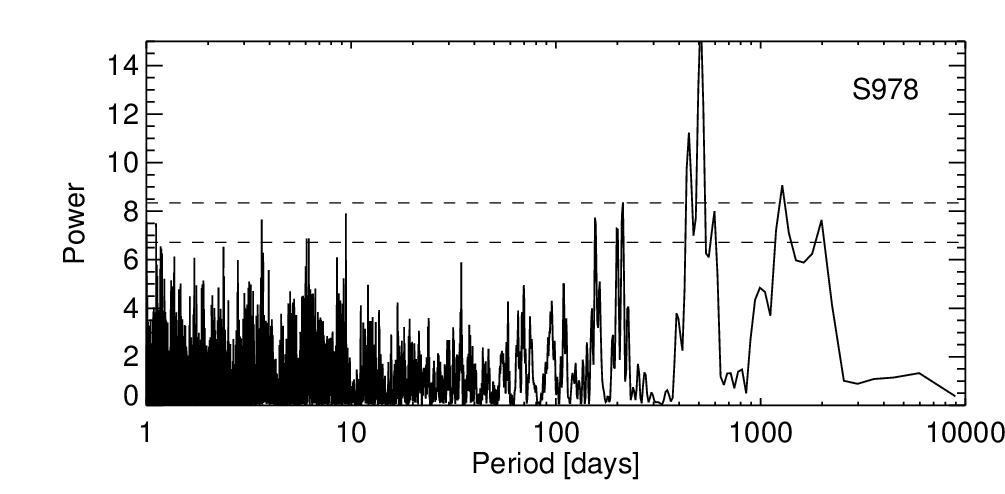} 
    \includegraphics{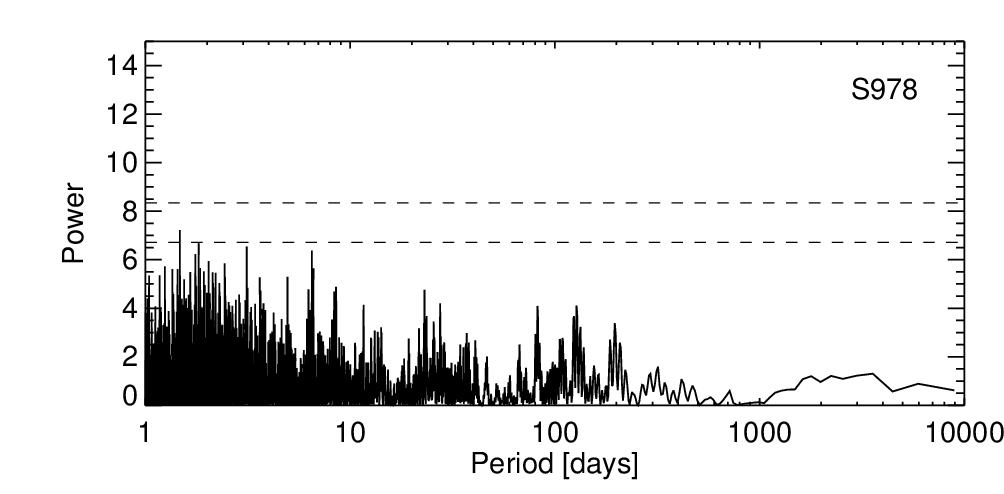} 
    \includegraphics{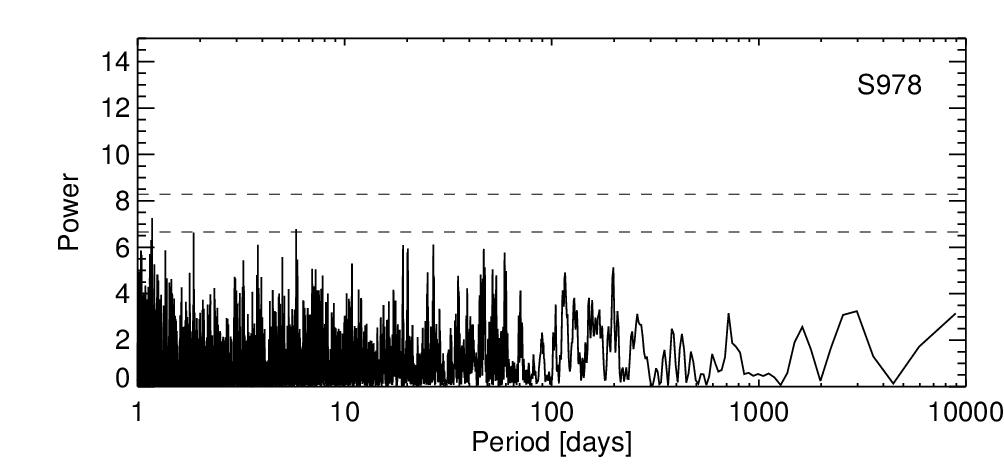}  
    \includegraphics{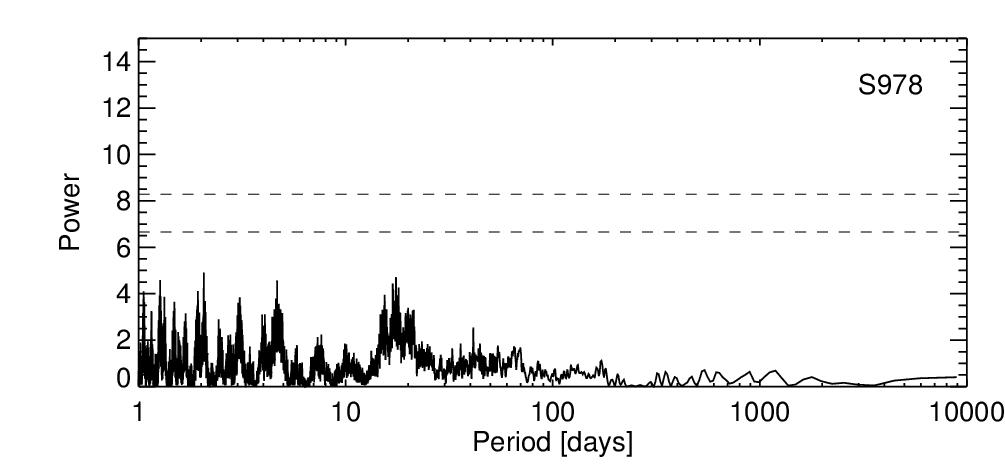}   
       } \\ 

                        \end{tabular}
                \end{center}
                \caption
           { \label{Periodograms} 
           Top: Lomb-Scargle periodogram of the RV measurements, residuals, bisector span, and FWHM
     for YBP778. Bottom: same plots for S978.  
     The dashed lines correspond to 5\% and 1\% FAPs, calculated according 
     to Horne \& Baliunas (1986) and white noise simulations.
                }
        \end{figure*}

  \begin{figure*}[h]
 \centering
 \resizebox{\hsize}{!}
            {
 \includegraphics{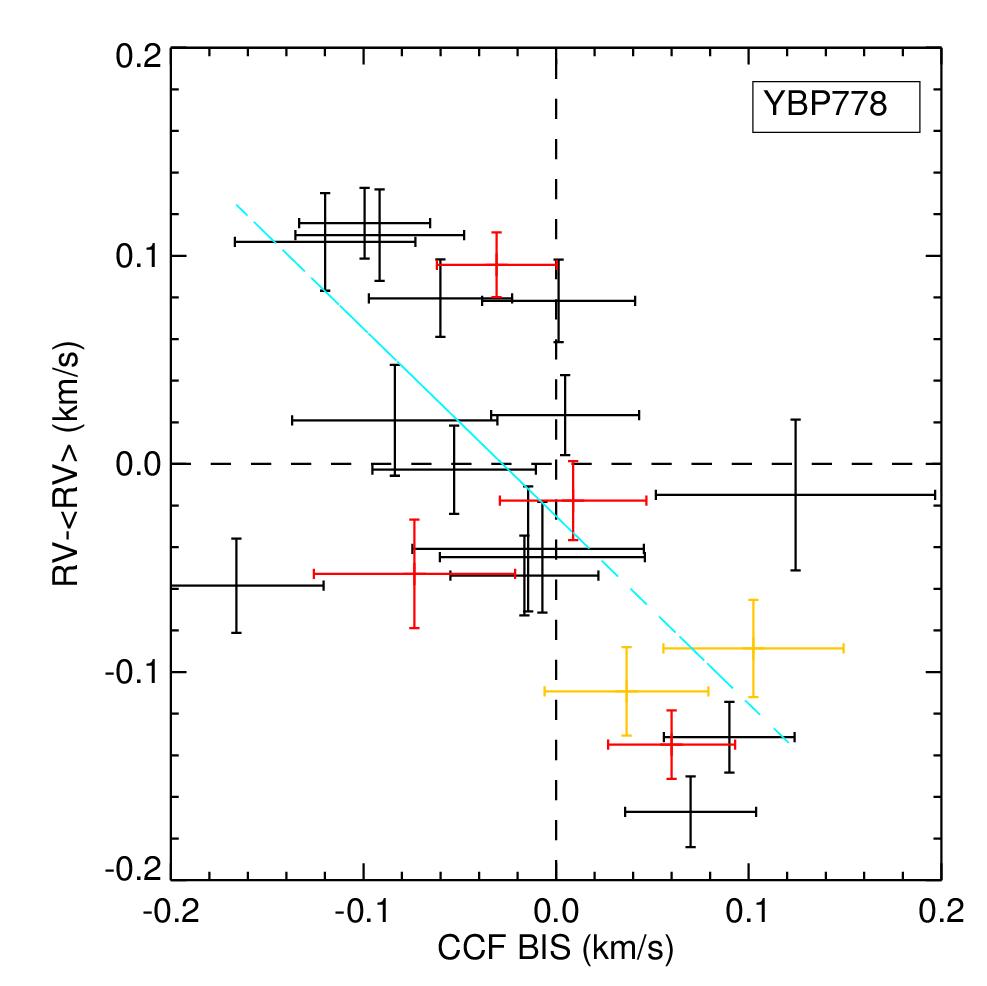}
 \includegraphics{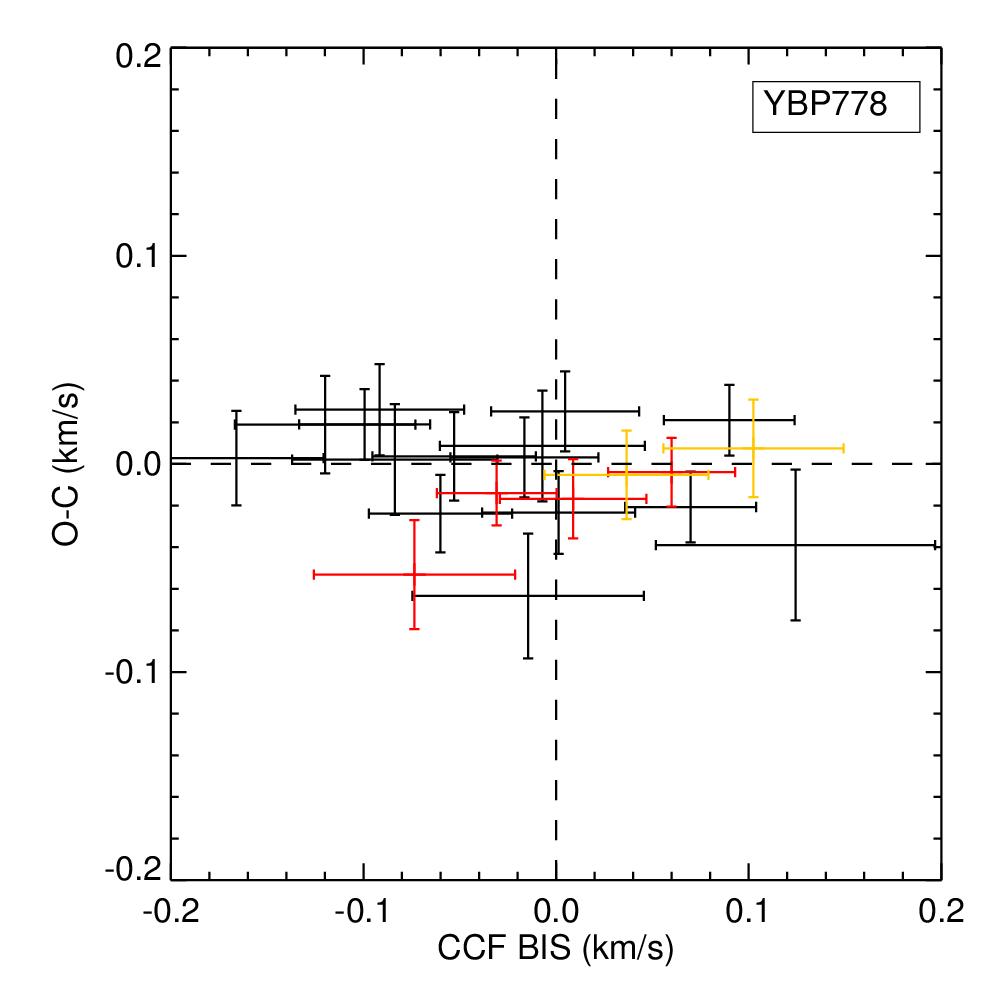}
 \includegraphics{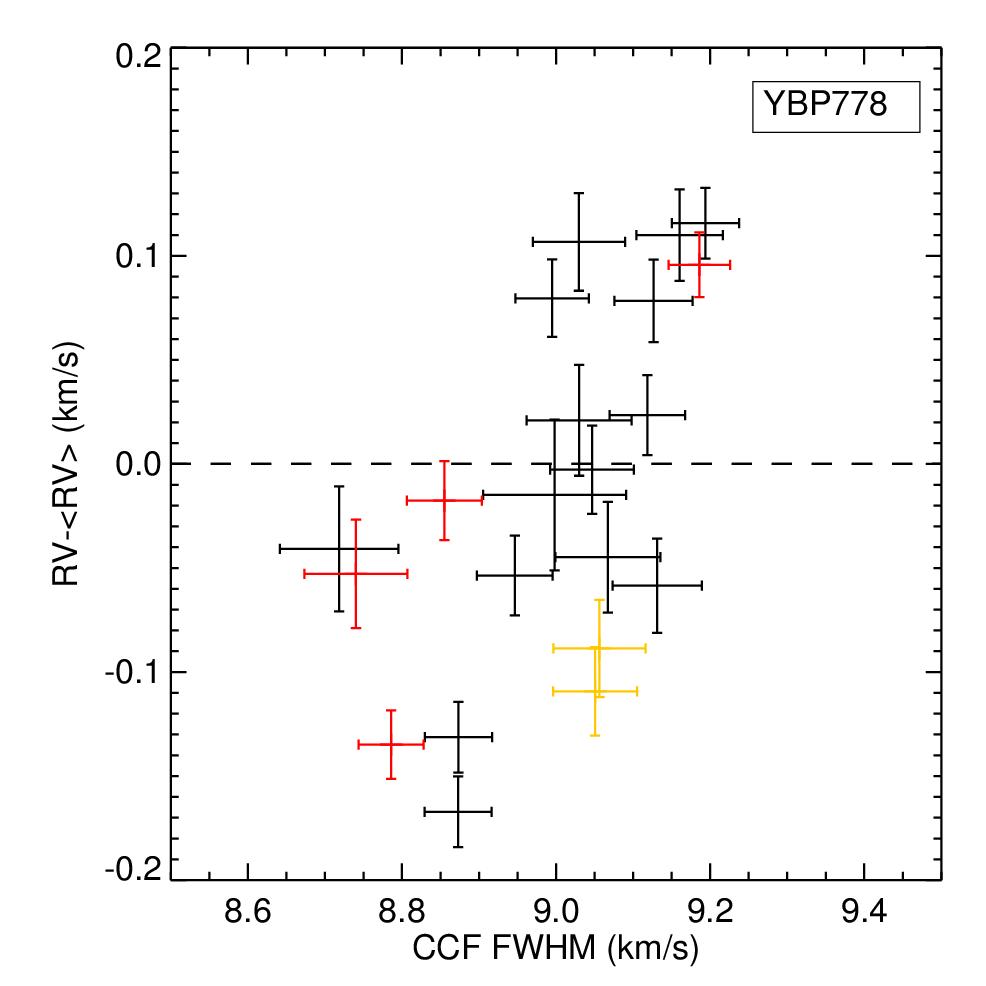}
 \includegraphics{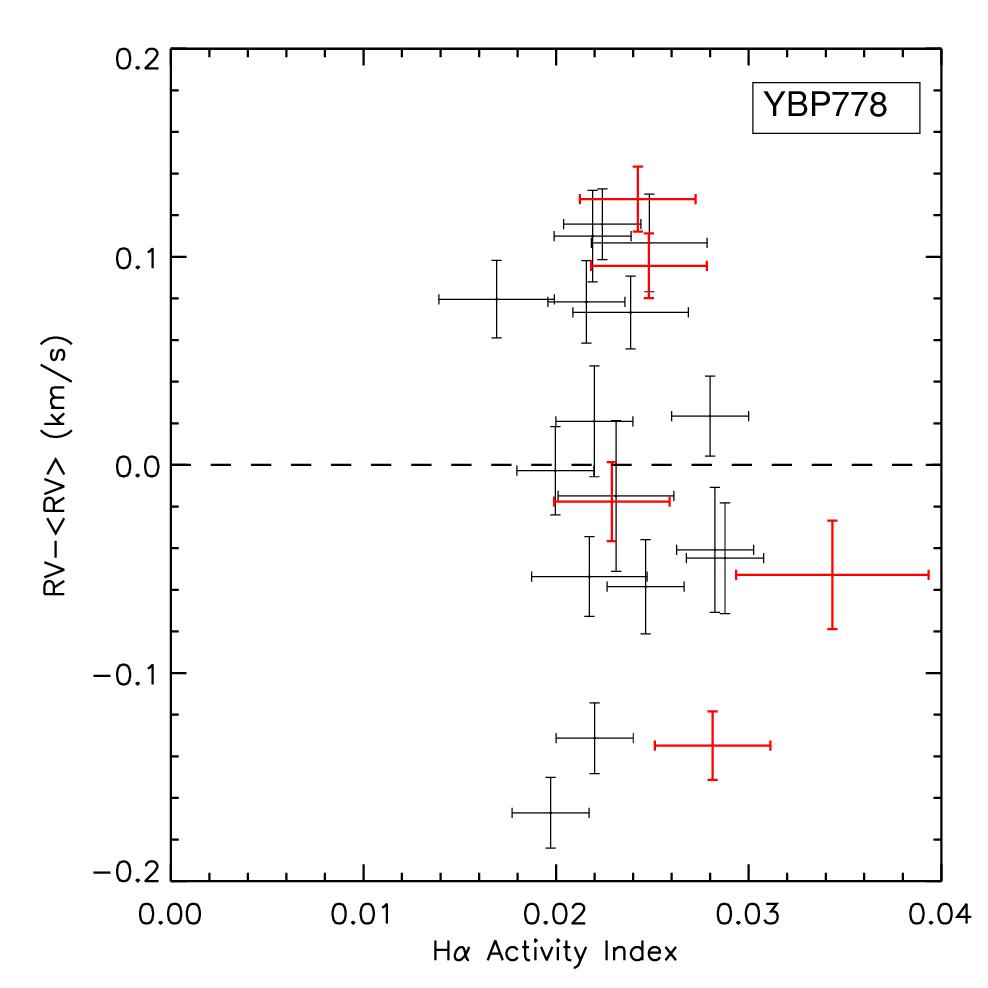}
  }
 \resizebox{\hsize}{!}
            {
 \includegraphics{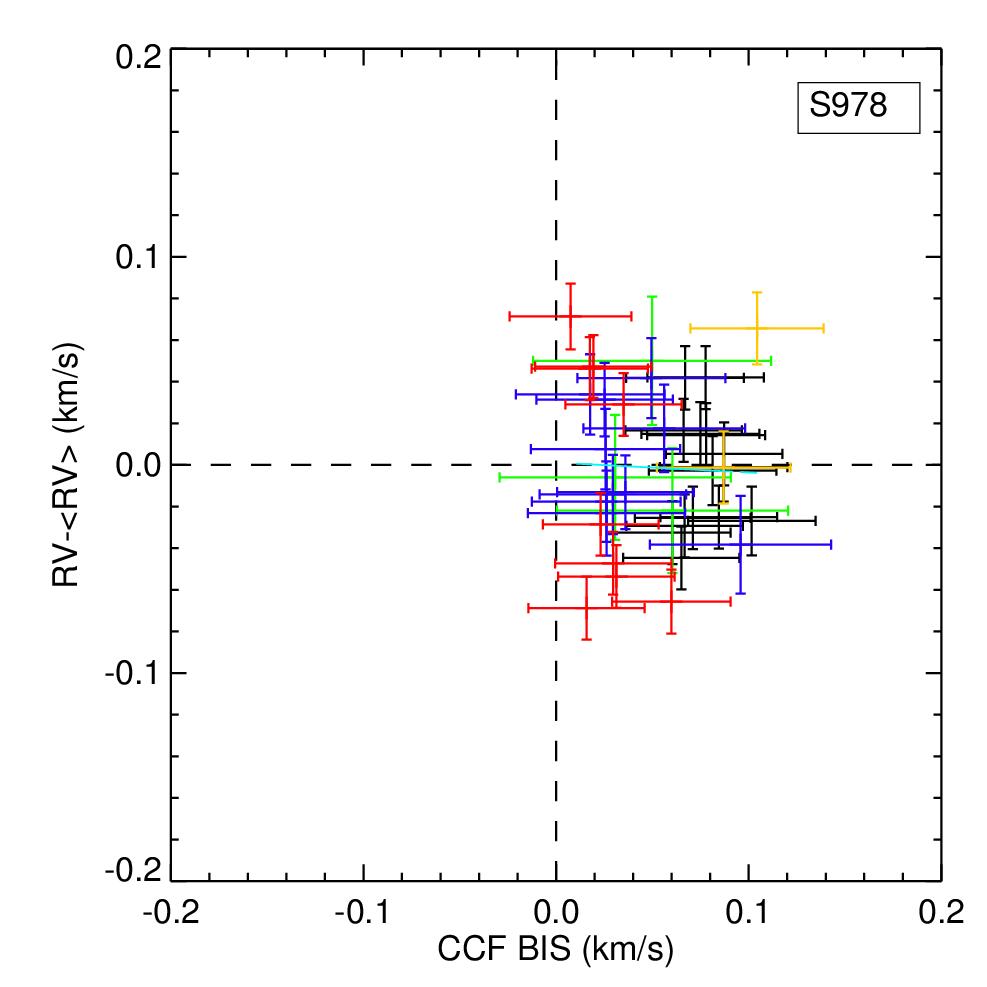}
 \includegraphics{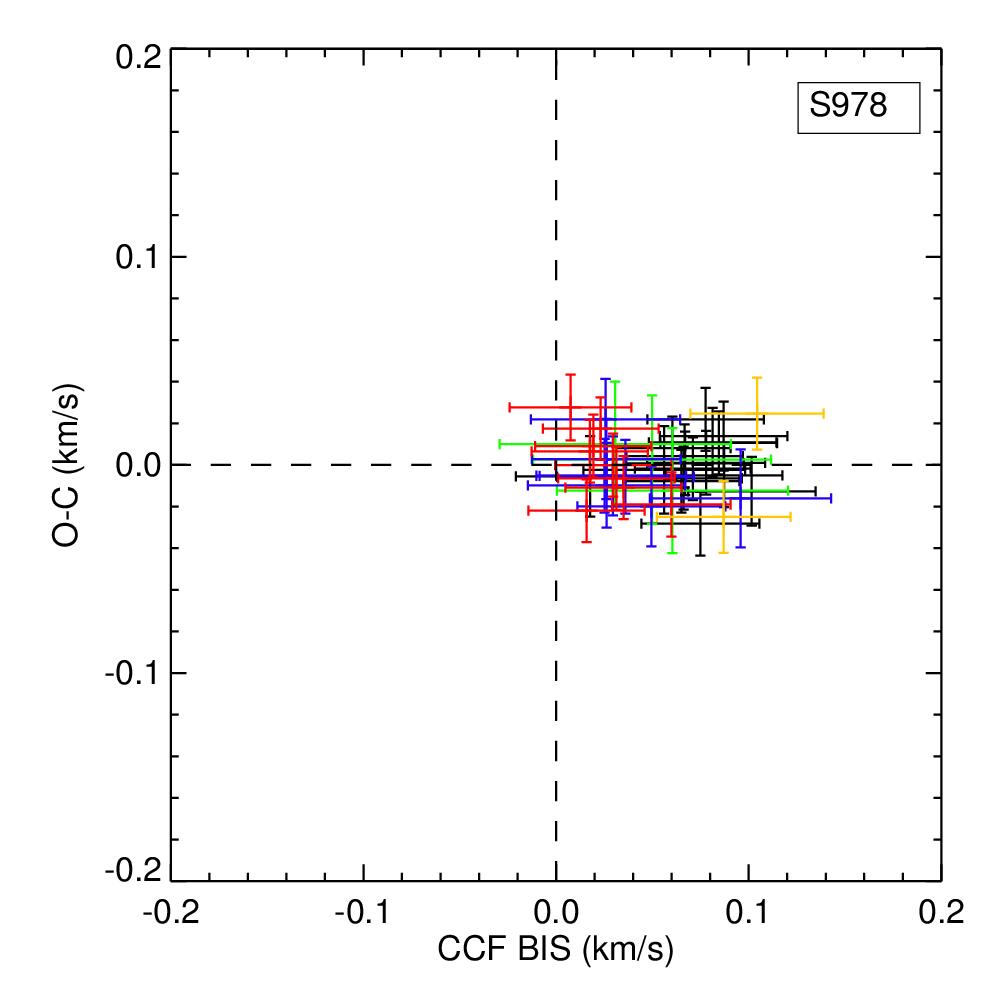}
 \includegraphics{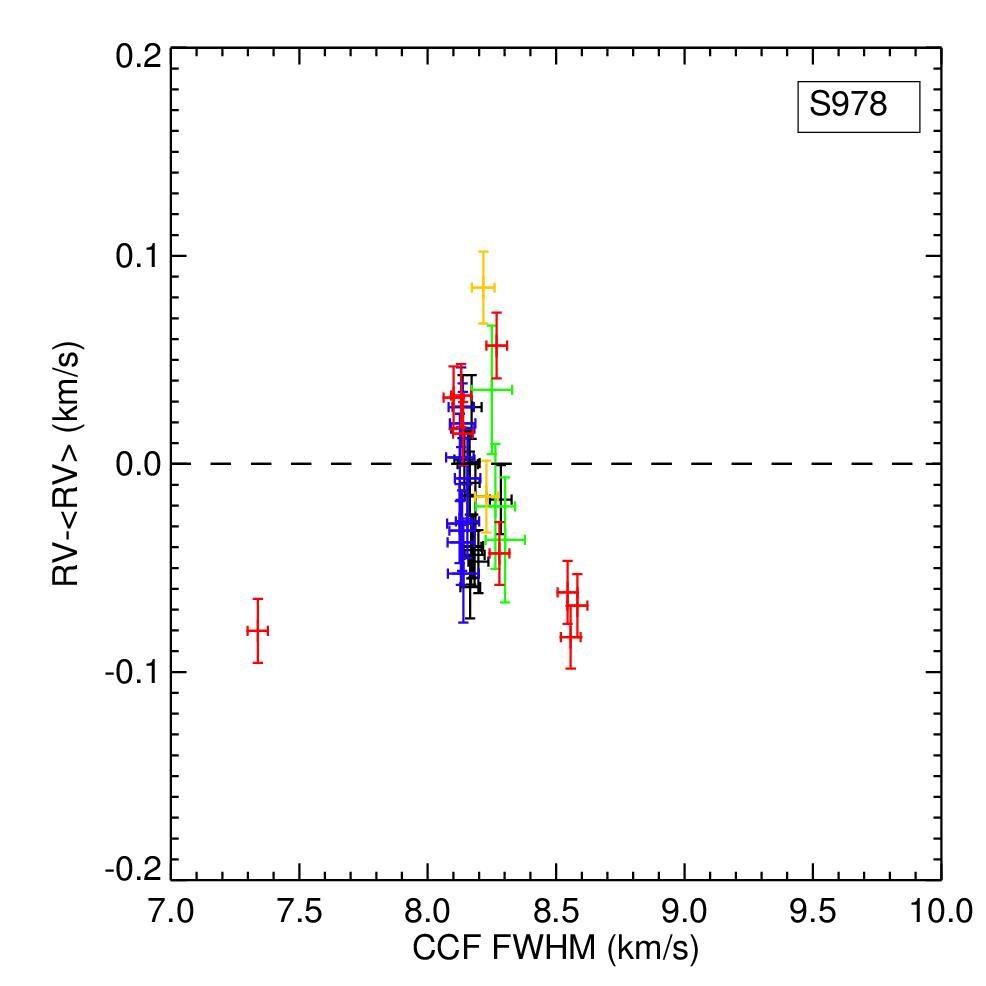}
 \includegraphics{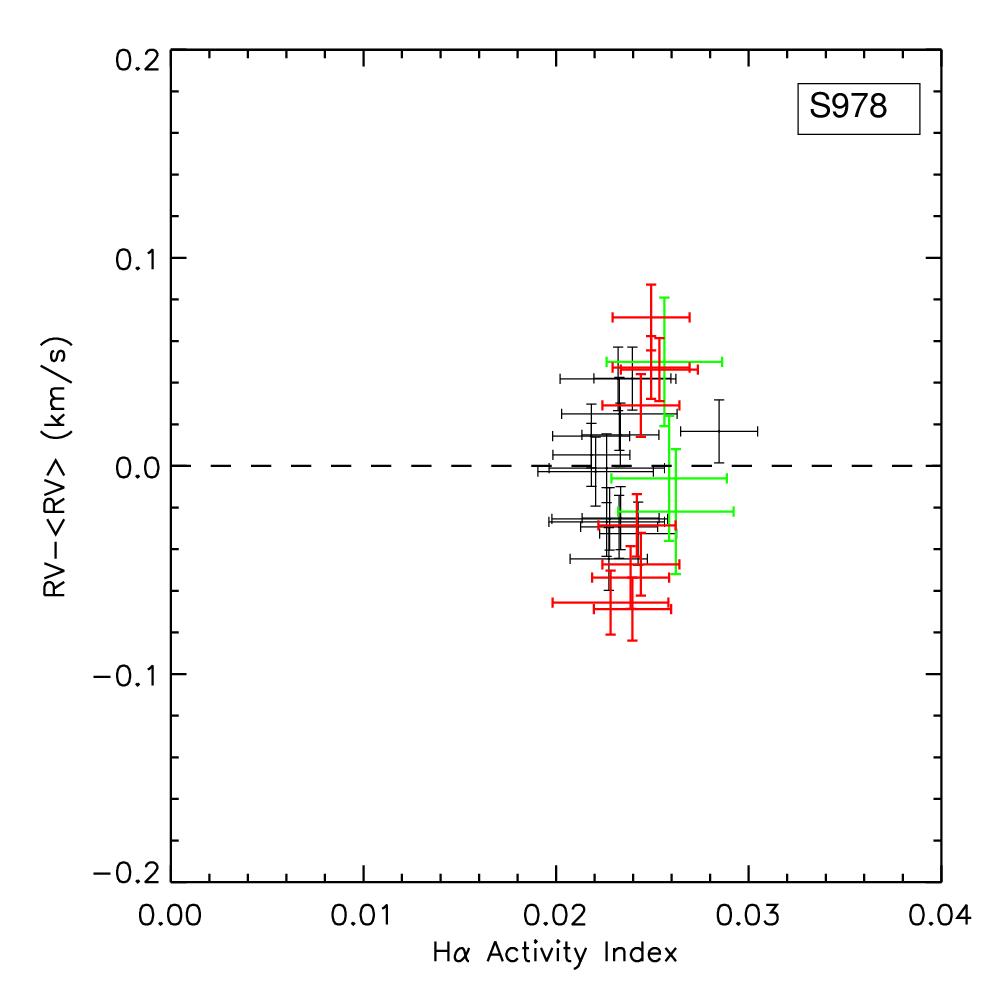}
  } 
  
 \caption{Top: RV measurements versus bisector span, residuals versus bisector span, RV measurements versus CCF FWHM, and 
 RV measurements versus H$\alpha$ activity indicator for YBP778. The H$\alpha$ activity indicator is computed as the area
 below the core of H$\alpha$ line with respect to the continuum. 
 CCF FWHM values are calculated by subtracting in quadrature the respective instrumental FWHM.
 The same symbols as in Fig.~\ref{Fit_YBP778}. 
 Bottom: The same plots for S978.
 }
 \label{BIS_FWHM_Ha}
 \end{figure*}

 \end{appendix}

\end{document}